\documentclass[useAMS,usenatbib]{mn2e}

\usepackage{times}
\usepackage{graphicx}
\usepackage{setspace}
\newcommand{\etal  }{{et al.} }
\newcommand{\msun}{\thinspace M_\odot}  
\newcommand{\vect}[1]{\mbox{\boldmath$#1$}}

\newcommand{\bzc}{ B_{\rm zc}}
\newcommand{\amt}{{ A_{\varphi}}}
\newcommand{\amz}{{ A_{z}}}
\newcommand{\rhoc}{\rho_{\rm c}}

\newcommand{\ob}{\varepsilon_{\rm ob}}
\newcommand{\ar}{\varepsilon_{\rm ar}}

\newcommand{\cm  }{\,{\rm cm}^{-3} } 
\newcommand{\fig}[1]{Fig.~\ref{fig:#1}}

\newcommand{\dfrac}[2]{{\displaystyle \frac{#1}{#2}}  }
\newcommand{\km}{\, {\rm km}~{\rm s}^{-1}}

\title[Collapse and Fragmentation of Molecular Clouds I]
{
Collapse and Fragmentation of Rotating Magnetized Clouds. 
I.\\ Magnetic Flux - Spin Relation
}
\author[M. ~N. ~Machida, T.~Matsumoto, K. ~Tomisaka and T. ~Hanawa]
{Masahiro ~N. ~Machida$^{1}$\thanks{E-mail:machida@cfs.chiba-u.ac.jp}, 
Tomoaki ~Matsumoto$^{2}$\thanks{E-mail:matsu@i.hosei.ac.jp} 
Kohji ~Tomisaka$^{3}$\thanks{E-mail:tomisaka@th.nao.ac.jp}
\newauthor and Tomoyuki ~Hanawa$^{1}$\thanks{E-mail:hanawa@cfs.chiba-u.ac.jp} 
\\
$^{1}$ Center for Frontier Science, Chiba University, Yayoicho 1-33, Inageku, Chiba 263-8522, Japan \\
$^{2}$ Faculty of Humanity and Environment, Hosei University, Fujimi, Chiyoda-ku, Tokyo 102-8160, Japan \\
$^{3}$ National Astronomical Observatory of Japan, Mitaka, Tokyo 181-8588, Japan
 }

\begin{document}



\maketitle

\begin{abstract}
We discuss evolution of the magnetic flux density and angular velocity in a molecular cloud core, on the basis of three-dimensional numerical simulations, in which a rotating magnetized
cloud fragments and collapses to form a very dense optically thick core of $ > \, 5 \times 10 ^{10} \cm $.
As the density increases towards the formation of the optically
    thick core, the magnetic flux density and angular velocity converge
    towards a single relationship between the two quantities.  
If the core is magnetically dominated its magnetic flux density approaches 
    $1.5 (n/ 5 \times 10^{10} \, \cm )^{1/2}$~mG, while if the core
    is rotationally dominated the angular velocity approaches $2.57 \times
    10^{-3} \, (n/5 \times 10^{10} \, \cm )^{1/2}$ yr$^{-1}$, where
    $n$ is the density of the gas.
We also find that the ratio of the angular velocity to
the magnetic flux density remains nearly constant
until the density exceeds $ 5 \times 10 ^{10} \cm $.
Fragmentation of the very dense core and emergence
of outflows from fragments are shown in the subsequent paper.
\end{abstract}


\section{Introduction}

It has long been recognized that magnetic field and 
rotation affect collapse of a molecular cloud, and accordingly, 
star formation.  The magnetic and centrifugal forces, 
as well as the pressure force, oppose the self-gravity of
the cloud and delay star formation.
Magnetic field and rotation are coupled.  Magnetic
field is twisted and amplified by rotation.  The twisted
magnetic field  brakes cloud rotation and launches outflows.  

In spite of its importance, only a limited number of 
numerical simulations have been performed for the coupling 
of magnetic field and rotation in a collapsing molecular cloud. 
The first numerical simulation of self-gravitating rotating magnetized 
clouds were performed by \citet{dorfi82}.  He found formation of 
bar-like structure for a cloud rotating perpendicular to the magnetic
field and that of ring-like structure for an aligned rotator.
However, the grid resolution was limited so that the simulation was stopped
when the density increased by 200 times from the initial value.
The spatial resolution was limited also in other simulations in 1980's by
\citet{phillips85} and \citet{dorfi89}, who studied the cloud 
with toroidal magnetic field and that with oblique magnetic field, 
respectively.  The spatial resolution was improved greatly by
\citet{tomisaka98,tomisaka02}.  He considered an initially filamentary cloud
of $ n _{\rm max} \, = \, 10 ^4 \, \cm $ and followed the evolution up to
the emergence of magnetically driven outflows from the first core of
$ n _{\rm max} \, > \, 10 ^{11} \, \cm $, where $ n _{\rm max} $ denotes
the maximum density.
However, his computation was two dimensional and could not take account of 
asymmetry around the rotation axis.
\citet{basu94,basu95a,basu95b} and 
\citet{nakamura02,nakamura03,li02}  have
got rid of the symmetry around the axis
but introduced the thin disk approximation.
The magnetic braking could not be taken into account in 
these simulations because of the thin disk approximation.
Although \citet{boss02} has performed three-dimensional simulations, 
he has employed approximate magnetohydrodynamical equations.
The approximation neglects torsion of the magnetic field
and replaces magnetic tension with the dilution of 
the gravity.  A fully three-dimensional numerical simulation
has just been initiated by \citet{machida04}, \citet{hosking04},
and \citet{matsu04}.  

The recent fully three-dimensional simulations have demonstrated
that fragmentation of the cloud depends on the magnetic field 
strength.  When the magnetic field is weak, a rotating cloud
fragments after the central density exceeds the critical
density, $ 5 \times 10 ^{10} \cm $, i.e., after the formation
of Larson's first core \citep{larson69}.
The magnetic field changes its direction and  strength during the collapse of the cloud.
Thus it is important to study how strong a magnetic
field the first core has.  

In this and subsequent papers, we show 144 models in
which a filamentary cloud collapses to form a magnetized
rotating first core.  
All the models are constructed using the fully three-dimensional numerical simulation code
used in \citet[][hereafter MTM04]{machida04}.  
This paper shows the evolution by the first core formation stage, i.e.,
the stages before the maximum density reaches 
the critical density, $ 5 \times 10 ^{10} \cm $.
The later stages, i.e., fragmentation of the first core
and emergence of outflows, are shown in the subsequent paper
\citep[][hereafter PaperII]{machida04-3}.  

From analysis of 144 models, we find two variables
which characterize the evolution of magnetic field
and rotation.  The first one is the ratio of the
angular velocity to the magnetic field.  This remains
nearly constant while the maximum density increases
from $ 5 \times 10 ^2 \cm $ to $ 5 \times 10 ^{10} \cm $.
The second characteristic variable is the sum of 
the ratio of the magnetic pressure to the gas pressure
and the square of the angular velocity in units of
the freefall timescale.  This variable converges to
a certain value.  We refer to the convergence as the magnetic
flux - spin ($ B - \Omega $) relation in the following.
We discuss the evolution of the magnetic flux density
and angular velocity by means of these two characteristic
variables.

This paper is organized as follows:  
Section 2 denotes the framework of our models and the assumptions
employed.  Section 3 describes methods of numerical simulations.
Section 4 presents typical models in the first four subsections
and compares various models in the last subsection.
Section 5 discusses implications of the magnetic flux - spin
relation and some applications of our models to observations.

\section{Model}

We consider formation of protostars through fragmentation of a filamentary molecular 
cloud by taking account of its magnetic field and self-gravity.  
The magnetic field is assumed to be coupled with the gas for simplicity although the molecular gas
is only partially ionized.  
Then the dynamics of the cloud are described by the ideal magnetohydrodynamical (MHD) equations,
\begin{eqnarray} 
& \dfrac{\partial \rho}{\partial t}  + \nabla \cdot (\rho \vect{v}) = 0, & \\
& \rho \dfrac{\partial \vect{v}}{\partial t} 
    + \rho(\mbox{\boldmath{v}}\cdot \nabla)\vect{v} =
    - \nabla P - \dfrac{1}{4 \pi} \vect{B} \times (\nabla \times \vect{B})
    - \rho \nabla \phi, & \\ 
& \dfrac{\partial \vect{B}}{\partial t} = 
   \nabla \times (\vect{v} \times \vect{B}), & \\
& \nabla^2 \phi = 4 \pi \rm{G} \rho, &
\end{eqnarray}
where $\rho$, $\vect{v}$, $P$, $\vect{B} $, and $\phi$ denote the density, 
velocity, pressure, magnetic flux density and gravitational potential, 
respectively. 
The ideal MHD approximation is fairly good as long as the gas
density is lower than $ \sim 10 ^{11} $~cm$^{-3}$ \citep{nakano88,nakano02}.  
The gas pressure is assumed to be
\begin{equation}
P = c_s^2 \rho \left[
1 + \left( \dfrac{n}{n_{\rm cri}}   \right)^{2/5} 
\right],
\label{eq:eos}
\end{equation}
where $n$ denotes the number density and is related to the mass density $\rho$ by 
\begin{equation}
\rho = 2.3 \times 1.67\times 10^{-24} \times n.
\end{equation}
The critical number density is set to be $n_{\rm cri} = 5 \times 10^{10} \cm$ \citep{masunaga00} and the sound speed is assumed to be $c_s = 0.19$~km~s$^{-1}$.
Thus, this equation of state means that the gas is isothermal at $T = 10$K for $n \ll n_{\rm cri}$ and adiabatic for $n \gg n_{\rm cri} $. 

Our initial model is the same as that of Tomisaka (2002) except for the azimuthal 
perturbation.  It is expressed as 
\begin{eqnarray}
\rho&=&\rho_{\rm c,0}\left[1+(r^2/8H^2) \right]^{-2}\left[1+\delta\rho_z(z)\right]
\left[1+\delta\rho_{\varphi}(r, \varphi)\right], 
\label{eq:petrho} \\
\vect{v} &=& r \, \Omega_{\rm c,0} \left[ 1 + (r^2/8 H^2)  \right]^{-1/2} \vect{e} _\varphi, 
\label{eq:v0} \\
\vect{B} &=& B_{\rm c,0} \left[ 1+ (r^2/8H^2) \right]^{-1}\left[1+\delta B_{z}(r,\varphi)\right] \vect{e} _z, 
\label{eq:petbz}
\end{eqnarray}
where
\begin{eqnarray}
 H^2 = \dfrac{c_s^2 +B_c^2/8 \pi \rho_c}{4 \pi G \rho_{\rm c,0} 
 - 2 \Omega_c^2} \ .
\label{eq:H}
\end{eqnarray}
in the cylindrical coordinates, $ (r, \, \varphi, \, z) $.
This initial model denotes a magnetohydrodynamical equilibrium 
\citep{sto63} when $ \delta \rho _z (z) $, 
$ \delta \rho _\varphi (r, \, \varphi) $ and $ \delta B _z (r, \varphi) $ are not taken into account.
The initial density is $ n _{\rm c, 0}  = 5 \times 10^2$~cm$^{-3}$ on
the axis ($ r \, = \, 0 $).
The filamentary cloud is supported in part by the magnetic field and rotation.
This equilibrium is unstable against fragmentation in the $ z $-direction.
The perturbation in the $ z $-direction
is assumed to be 
\begin{eqnarray}
\delta\rho_z &=&  \amz \cos 
\left( 2\pi z / \lambda _{\rm max} \right),
\end{eqnarray}
where
\begin{eqnarray}
\lambda_{\rm max}\simeq
 \left[  \dfrac{c_s}{(4\pi G \rho_{\rm c,0})^{1/2}}
 \right] 
\dfrac{2\pi (1+\alpha/2+\beta)^{1/2}}{0.72 \left[ (1 + \alpha/2 + \beta)^{1/3}-0.6 \right]} ,
\label{eq:lambda}
\end{eqnarray}
and
\begin{equation}
\beta \equiv 2 \omega_c^2 H^2/c_s^2 .
\end{equation}
The symbol, $ \lambda _{\rm max} $, denotes 
the wavelength of the fastest growing perturbation \citep{matsu94}.

The azimuthal perturbation is assumed to be
\begin{eqnarray}
\delta\rho_{\varphi},\delta B_{\varphi} &=& \left\{ 
\begin{array}{ll}
\amt(r/H)^m  \cos(m\varphi),  \hspace{5pt} \ \rm{for} \ \it{r_c\le H},  \\
\amt \cos( m \varphi), \ \hspace{36pt} \rm{for} \ \it{r_c>H},
\end{array} 
\right.  
\end{eqnarray}
where the azimuthal wavenumber is assumed to be $ m $ = 2.
The radial dependence is chosen so that the density perturbation
remains regular at the origin ($ r \, = \, 0 $) at one time step
after the initial stage.
The ratio of density to the magnetic flux density is constant in the
$ \varphi $-direction for a given $ r $ and $ z $ [see equations (\ref{eq:petrho})
and (\ref{eq:petbz}) ].

The initial model is characterized by four nondimensional parameters: 
twice the magnetic-to-thermal pressure ratio,
\begin{equation}
\alpha =  B_{\rm zc,0}^2 / (4\pi \rho_{\rm c,0} c_{\rm s,0}^2) ,
\label{eq:alpha}
\end{equation}
the angular velocity normalized by the free-fall timescale,
\begin{equation}
\omega = \Omega_{\rm c,0}/\sqrt{4 \pi {\rm G} \rho_{\rm c,0} } ,
\label{eq:omega}
\end{equation}
the amplitude of the perturbation in the $ z $-direction, $ \amz $,
and that of the non-axisymmetric perturbation, $ \amt $.
The former two specify the equilibrium model, while the later two
do the perturbations.
We made 144 models by combining values listed in Table~\ref{table:para}.
The results depend little on the values of $ \amz$, thus $ \amz $ is fixed to be 0.1 
in most models.

\section{Numerical Method}

We employed the same 3D MHD nested grid code as that used in
MTM04.  It incorporates the
3D nested grid code of \citet{matsu03b} for a hydrodynamical
simulation and the approximate Riemann solver for the MHD equation
\citep{fukuda99}.
This 3D MHD nested grid code integrates equations (1) through (5) numerically, 
with a finite difference scheme on the Cartesian coordinates.
The solution is a second order accurate, both in space and in time by virtue of the Monotone Upstream Scheme for Conservation Law (see e.g., Hirsh 1990).
The Poisson equation is solved by the multigrid iteration \citep{matsu03a}.
We have used Fujitsu VPP 5000, vector-parallel supercomputers, 
for 40 hours to make a typical model shown in this paper.

The nested grid consists of concentric hierarchical rectangular subgrids
to gain high spatial resolution near the origin.
Each rectangular grid has the same cell number ($ = 128 \times 128
\times 32 $) but a different 
cell width, $ h (\ell) \, \equiv \, 2 ^{-\ell-5} \,
\lambda _{\rm max} $, where $ \ell $ denotes the level of the grid
and ranges from 1 to $ \ell _{\rm max} $.
Thus the coarsest rectangular grid of
$ \ell \, = \, 1 $ covers the whole computation region
of $ - \lambda _{\rm max} \, \le \, x \, \le \lambda _{\rm max} $,
$ - \lambda _{\rm max} \, \le \, y \le \lambda _{\rm max} $, and
$ 0 \, \le \, z \le \lambda _{\rm max} /2 $.  The solution in
$  z \, < \, 0 $ is constructed from that in $ z \, \ge \, 0 $
by the mirror symmetry with respect to $ z \, = \, 0 $.
The maximum level number is set at $ \ell _{\rm max} \, = \, 3 $
at the initial stage ($ t \, = \, 0 $).  A new finer subgrid is 
generated whenever the minimum local Jeans length 
$ \lambda _{\rm J} $ becomes smaller than $ h (\rm \ell_{\rm max}) / 8 $.
Since the density is highest always in the finest subgrid, the generation
of the new subgrid ensures the Jeans condition with a margin of 
a factor of 2 \citep[c.f.][]{truelove97}.
We have adopted the hyperbolic divergence cleaning method
of \citet{dedner02} to obtain the magnetic field of $ 
\mbox{\boldmath$\nabla$}\cdot\mbox{\boldmath$B$} $ free.

\section{Results}
\label{sec:results1}

We have followed all the models shown in this paper until
the central density exceeds $ n _{\rm c} \, \ga \, 10 ^{15} \cm $.
This paper describes the first half of the evolution for
each model, i.e., the stages of
$ n _{\rm c} \, \le \, n _{\rm cr} \, = \, 
5 \times 10^{10} \cm $.  The second half
is described in the subsequent paper (Paper II).


Our models are characterized mainly by the strength of 
the magnetic field ($\alpha$) and the angular velocity ($ \omega $).  
They are classified into four groups: (A) models having small 
$ \alpha \, (< 0.1) $ and small $ \omega \, (<0.1) $, 
(B) those having large $ \alpha (\ge 0.1) $ 
and small $ \omega \, (< 0.1) $, 
(C) those having small $ \alpha \, (< 0.1) $ and large 
$ \omega \, (\ge 0.1 ) $, and 
(D) those having large $ \alpha \, (\ge 0.1) $ and large  
$\omega \, (\ge 0.1) $.  In other words, the model cloud has 
a weak magnetic field and rotates slowly in group A, while 
it has a relatively strong magnetic field and relatively 
large angular momentum in group D.  
Each group is described separately in the following subsections, 
in each of which two typical models of $ \amt \, = \, 0.01 $ (S) 
and 0.2 (L) are shown.  
The typical models are named after the group (A, B, C, or D) and 
$ \amt $ (S or L).
Model AS has $ \alpha \, = \, 0.01 $, $ \omega $ = 0.01, and $ \amt $ = 0.01, for example.
Table~\ref{table:init} shows the values of $ \alpha $, $ \omega $, $ \amz $, and $ \amt $  for the 8 typical models shown in the following subsections.
It also shows the initial magnetic field ($ B _{zc,0} $), the initial angular velocity ($ \Omega _{c0} $), the wavelength of the perturbation in the $ z $-direction ($ \lambda _{\rm max} $),  the mass ($ M $) of the gas contained in the region of $ \vert z \vert \, \le \, \lambda _{\rm max} / 2 $, and the epoch at which the density becomes infinity ($t_{\rm f}$).

\subsection{Weak Magnetized and Slowly Rotating Cloud}
\label{sec:modelA}

\begin{figure}
\begin{center}
\includegraphics[width=160mm]{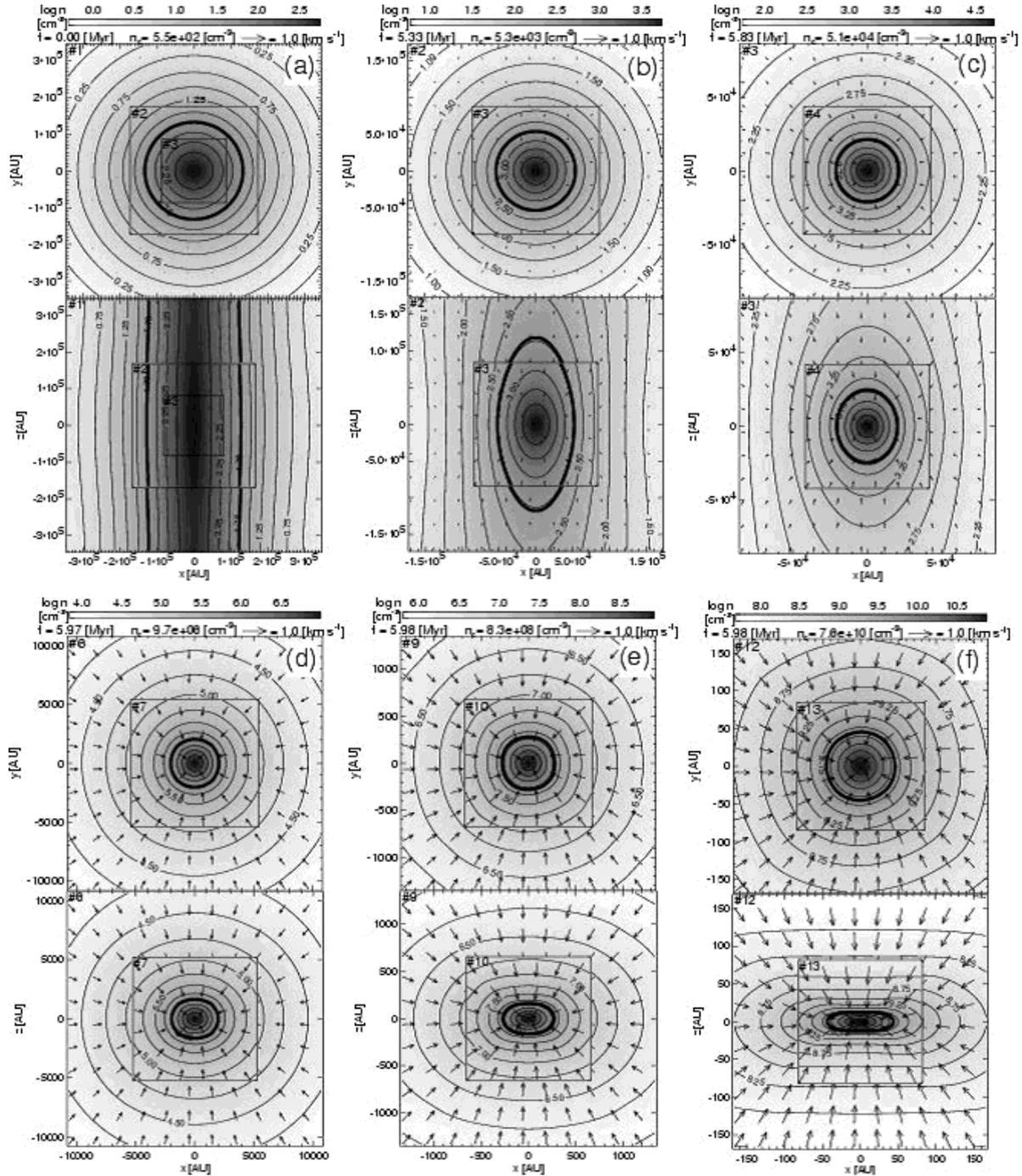}
\caption{
The density (grey-scale and contour) and velocity distributions (arrows) in model AS [($\alpha$, $\omega$, $\amt$)=(0.01, 0.01, 0.01)].
Panels (a) through (f) are snapshots at the stages of 
(a) $n_c =5.5 \times 10^2 \cm$ ($l=1, 2$), \ 
(b) $ 5.3 \times 10^3 \cm$ ($l=2, 3$), \ 
(c) $ 5.1 \times 10^4 \cm$ ($l=3, 4$), \
(d) $ 9.7 \times 10^6 \cm$ ($l=6, 7$), \
(e) $ 8.3 \times 10^8 \cm$($l=9, 10$), \ and \
(f) $ 7.6 \times 10^{10} \cm$($l=12, 13$),
where $l$ denotes the level of subgrid.
The level of the subgrid is shown in the upper left corner of each solid square which denotes the outer boundary of the subgrid. 
The upper and lower panels show the cross sections at $z$=0 and $y$=0 planes, respectively. 
The grey-scale is different for each panel as shown above the panels.
Thick contours denote the density of  $n = 1/10\, n_c$, where $n_c$ means density at the center.
The elapsed time, density at the center and arrow scale are denoted in each panel.
}
\label{fig:A1}
\end{center}
\end{figure}

This subsection displays model AS as a typical model having 
a weak magnetic field and slow rotation.
Model AS has parameters $\alpha=0.01$, $\omega=0.01$ and $\amt=0.01$.
Fig.~\ref{fig:A1} shows the cloud evolution in model AS 
by a series of cross sections.

As shown in Fig.~\ref{fig:A1}, a gas cloud is transformed from a prolate
one to an oblate one on the $y=0$ plane (see lower panels), while it maintains 
a round shape on the $z=0$ plane (upper panels) in the period of 
$5.5 \times 10^2 \cm < n_c < 7.6 \times 10^{10} \cm$. 
The velocity field is almost spherically symmetric while the central density increases
 from $ 5\times 10^5 \cm $ to $2\times 10^9 \cm$.
The dense cloud is prolate and elongated in the $ z $-direction 
in the lower panel of \fig{A1} (b),  while it is nearly spherical 
in the lower panel of \fig{A1} (c).
In this early collapse phase, the cloud contracts along the major axis (i.e. $z$-axis), regardless of the magnetic field  and rotation as discussed in \citet{bonnell96}.  
An oblate core is seen in panel \fig{A1} (e) and a thin disk is 
seen in the lower panel of \fig {A1} (f). 
The collapse is dynamical at the stages shown in \fig{A1} (c) through 
\fig{A1} (f).  
The radial infall velocity reaches $ v _r $ = $ - 0.52 $~km~s$^{-1}$ on the 
$ z \, = \, 0 $ plane in \fig{A1} (e), while the vertical infall velocity does 
$ v _z \, = \, \pm 0.58 $~km~s$^{-1}$ on the $ z $-axis.  
The rotation velocity is $ v _\varphi $ = 0.047~km~s$^{-1}$ at 
maximum and much smaller than the infall velocities.  
This means gas contracts spherically in this phase.
The difference between the radial and vertical infall 
velocities is still small 
($ \vert v _{r,{\rm max}} \vert \, = \, 0.61 $ km~s$^{-1}$ and $ 
\vert v _{z,{\rm max}} \vert \, = \, 0.8 $ km~s$^{-1}$) in 
\fig{A1} (f) although a high density disk is formed.  

\begin{figure} 
\begin{center}
\includegraphics[width=120mm]{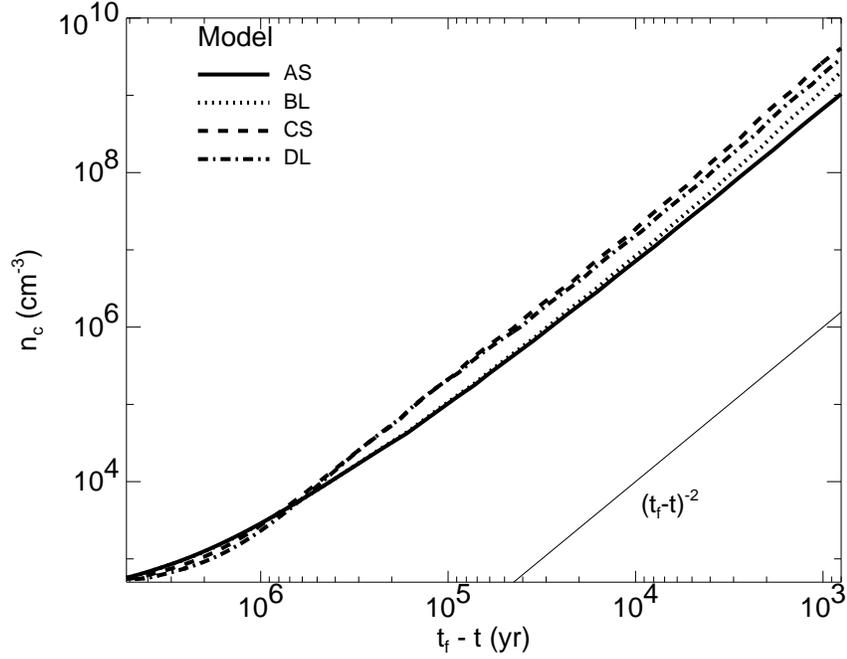}
\caption{
The central densities are shown as a function of time for models AS, BL, CS, and DL.
The value of $t_{\rm f}$ is shown in Table~\ref{table:init}.
See text for the definition of $t_{\rm f}$.
The relation of $n_c \propto (t-t_{\rm f})^{-2}$ is also plotted for comparison.  
}
\label{fig:dens}
\end{center}
\end{figure}

The density increase is well approximated by $ \rho _{\rm c} \, = \,
2.2 / [4 \pi G \, (t \, - \, t _{\rm f}) ^2] $ in the period of 
$ 5\times 10^4 \cm \, \le \, n \, \le \, 10^{9} \cm $ as shown
by the thick solid curve in Fig.~\ref{fig:dens}.
The offset is taken to be $ t _{\rm f} \, = \, 5.96 \times 10^6 $yr so that
the central density increases in proportion to the inverse 
square of the time in the widest span in $ \log \rho _{\rm c} $, 
as shown in \citet{larson69}. 
Remember that the similarity solution of Larson (1969) and Penston (1969) gives 
$ \rho _{\rm c} \, = \, 1.667 / [4 \pi G \, (t \, - \, t _{\rm f}) ^2] $
for spherical collapse of a non-magnetized non-rotating isothermal
cloud.  
We have checked that the density increase is well approximated by $ \rho _{\rm c} \, \simeq \,1.65 / [4 \pi G \, (t \, - \, t _{\rm f}) ^2] $ in a non-magnetized and non-rotating cloud of our  test calculation.
The density increase is 15~\% slower in model AS than in
the similarity solution, since $ (2.2/1.667)^{1/2} \, \simeq \, 1.15 $.  
This small difference is due to the 
rotation and magnetic field.

To evaluate the change in the core shape shown in \fig{A1},
we measure the moment of inertia for the high density gas of
$ \rho \, \ge \, 0.1  \rho _{\rm c} $.
We derive the major axis ($ h _l $), minor axis ($ h _s $), 
and $ z $-axis ($ h _z$) from
the moment of inertia according to \citet{matsu99}.
The oblateness is defined as $ \ob \, \equiv \, 
(h _l h _s) ^{1/2} / h _z $ and the axis ratio is 
defined as $ \ar \, = \, h _l / h _s \, - \, 1 $.

\begin{figure}
\begin{center}
\includegraphics[width=120mm]{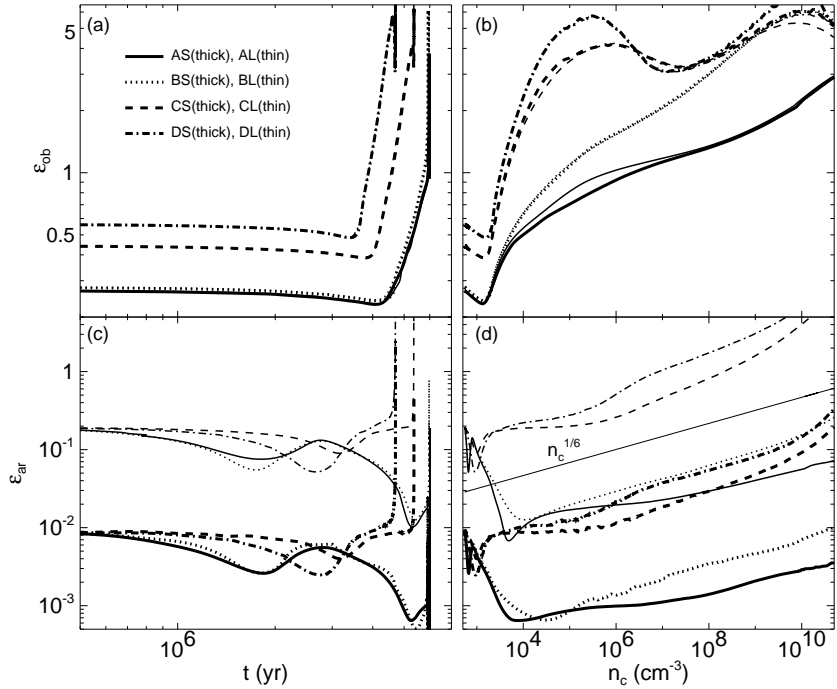}
\caption{
The oblateness (upper panels) and axis ratio (lower panels) are plotted against the elapsed time (left panels) and central density (right panels) for models AS, AL, BS, BL, CS, CL, DS and DL.
The relation of $\ar \propto n_c^{1/6} $ is also plotted in panel (d) for comparison.
}
\label{fig:ob}
\end{center}
\end{figure}

The oblateness is denoted by the thick solid curve as a function
of time in Fig.~\ref{fig:ob} (a) and as a function of 
the central density in Fig.~\ref{fig:ob} (b). 
The oblateness is nearly constant at $ \ob = 0.27 $ in the
 period of $t\ \la 4 \times 10^6$yr 
(or $5 \times 10^2 \cm \la n_c \la 5\times 10^3 \cm$).  
It increases and reaches $ \ob = 1 $ at the stage of
$ n _c = 2 \times 10 ^6 \cm $, which is shown in
the lower panel of Fig.~\ref{fig:A1} (c).  
The oblateness reaches  $\ob = 2.9$, 
when the disk-like structure is formed 
at $n_c = 7.6 \times 10^{10} \cm$ as shown 
in the lower panel of \fig{A1} (f).
The increase in $ \ob $ is monotonic in 
the period of $ n _{\rm c} > 5 \times 10 ^3 \cm $.

The axis ratio is denoted by the thick solid curve as a function of time in Fig.~\ref{fig:ob} (c) and as a function of the central density in Fig.~\ref{fig:ob} (d).  
The axis ratio decreases from $ \varepsilon _{\rm ar} $ = 0.01 to 
$ 7 \times 10 ^{-4} $ after oscillating once over the period
of $t \le 5\times 10^6$ yr ( or $ n _{\rm c} < 5 \times 10 ^3 \cm $).
It increases in proportion to $ n _{\rm c} ^{1/6} $ over the
period of $ n _{\rm c} > 5 \times 10 ^3 \cm $.
The growth rate of $ \varepsilon _{\rm ar} $ coincides with that of the
bar mode growing in the spherical runaway collapse (Hanawa \& Matsumoto 1999).
The axis ratio grows up to $3.5 \times 10^{-3}$ 
in the isothermal collapse phase as shown in \fig{ob} (d).
In order to examine  dependence on the axis ratio, 
we compare models AS and AL, of which model parameters are the
same except for the amplitude of the non-axisymmetric perturbation,
$ \amt $. The value of $ \varepsilon _{\rm ar} $ is 20 times 
larger in model AL than
in model AS at a given stage and reaches $\ar = 6.8 \times 10^{-2}$ 
at $ n _{\rm c} \, = \, n _{\rm cri} $.  The axis ratio is 
proportional to $ \amt $.
The oblateness is nearly the same in models AS and AL.  
The non-axisymmetric perturbation grows linearly in models AS and
AL.

The magnetic flux density increases as 
the density increases.  The left panel of \fig{bz} shows 
the square root
of the ratio of the magnetic pressure to the gas pressure,
$ \bzc / (8 \pi \rhoc c _{\rm s} ^2) ^{1/2} $, as a
function of $ n _{\rm c} $.  
Note that the ordinate is normalized by the initial value.
It increases in proportion to one sixth the power of the
density, i.e., 
$ \bzc / (8 \pi \rhoc c _{\rm s} ^2) ^{1/2} \propto
n _{\rm c} ^{1/6} $, in the period of
$ 10 ^6 \cm \leq n _{\rm c} \leq 
10 ^9 \cm $.  This means that the magnetic
field increases in proportion to $ \bzc \, \propto \,
n _{\rm c} ^{2/3} $.  This increase in $ \bzc $ is 
consistent with the spherical collapse of the core.
When the collapse is spherically symmetric, the
density and magnetic field increase inversely
proportional to the cubic and square of the radius,
respectively, since the magnetic field is frozen in
the gas.  
Hence, the magnetic field is proportional
to two thirds the power of the density,
$ \bzc \propto \rhoc ^{2/3} $. 

After the central density exceeds $ 10 ^9 \cm $, the growth of the magnetic field slows down.
This slowdown coincides with the change in the core shape.
The core is significantly oblate in the period of $ n _{\rm c} > 10 ^9 \cm $.  
Remember that the magnetic field is proportional to the square root of the density ($ \bzc \propto n_c ^{1/2} $) when a magnetized disklike gas cloud collapses (Scott \& Black 1980).
This is because the disk is nearly in a hydrostatic equilibrium in the $z$-direction and the isothermal disk has the relation $n \propto \Sigma^2$.
We use the terminology, the ``disk collapse", for this radial
collapse of a disklike gas cloud.
In the disk collapse, the magnetic flux density increases in proportion to 
the surface density ($ \bzc \propto \Sigma _{\rm c} $) 
since the gas is frozen in a magnetic flux tube. 
The relations, $ n_{\rm c} \propto \Sigma _{\rm c} ^2 $ and
$ \bzc \propto \Sigma _{\rm c} $, yield 
$ \bzc \propto n_{\rm c} ^{1/2} $.  
In the period of $ n _{\rm c} >  10 ^9 \cm $,
the growth rate of the magnetic field is intermediate
between those for the spherical collapse and for the
disk collapse.  This is consistent with the density
change over the same period.

As well as the magnetic flux density, the angular velocity 
of the core increases as the density increases.  
The right panel
of Fig. 4 shows the ratio of the angular velocity
to the magnetic field ($ \Omega _{\rm c} / \bzc $) normalized by the initial value 
 ($\Omega_{\rm c,0}/B_{zc,0}$)
as a function of $ n _{\rm c} $.  The ratio is 
nearly constant at the initial value.  This is because
both the specific angular momentum ($ j $) and the magnetic
flux ($ \Phi $) are conserved for a central magnetic
flux tube.  
Both the angular velocity and magnetic field increase proportionally to the inverse square of the tube radius.  
Hence the ratio is constant in both the spherical and  the disk collapse.  
The conservation of the specific angular momentum implies that none of 
the magnetic torque, gravitational torque, and 
$ \varphi $-component of the pressure force are significant.

Since $ \Omega _{\rm c} / \bzc $ is nearly constant, the angular velocity increases in proportion to $ n_{\rm c} ^{2/3} $ in the period of $ 10^6 \cm \leq n _{\rm c} \leq 10 ^9 \cm $
and the growth of $ \Omega _{\rm c} $ slows down in the period of $ n _{\rm c} > 10 ^9 \cm $.
When measured in units of the free fall timescale, the angular velocity increases in proportion to $ \Omega _{\rm c} (4 \pi G \rhoc) ^{-1/2} \propto n_{\rm c} ^{1/6} $ in the former period.
The angular velocity in units of the free fall timescale denotes
the square root of the ratio of the
centrifugal force to the gravitational force.
The magnetic field and rotation strengthen in
the same manner during the spherical collapse, since
both $ \bzc (8 \pi \rhoc c _{\rm s} ^2) ^{-1/2} $
and $ \Omega _{\rm c} (4 \pi G \rhoc) ^{-1/2} $
increase in proportion to $ \rhoc ^{1/6} $.

Model AS is similar to model B of \citet{matsu97}, 
although our model AS includes a very weak magnetic field.
The magnetic field influences little the cloud collapse.
\fig{A2} shows that the magnetic field is not twisted but 
pinched at the stages of $ n _{\rm c} \, \ge \, 10 ^6 $ cm$^{-3}$ 
as shown in panels (d) - (f) of \fig{A2}.
Each panel denotes the magnetic field lines for the corresponding
stage shown in each panel of \fig{A1}.  This weak magnetic
field has no significant effect.  
When $ \alpha \, < \, 0.1 $ and $ \omega \, < 0.1 $, 
the effects of magnetic field and rotation are very small.

\subsection{Strongly Magnetized and Slowly Rotating Cloud}
\label{sec:modelB}

Model BL is shown as a typical example of models in this subsection having large 
$ \alpha $ and small $ \omega $.
Model BL has  parameters  $\alpha=0.1$, $\omega=0.01$ and $\amt=0.2$.
The parameters of model BL are the same as those of 
model AL except for $ \alpha $,  which is 0.01 for model AL 
and 0.1 for BL (Table~\ref{table:init}).
When $ \alpha \, \ge \, 0.1 $, the magnetic pressure becomes
comparable to the gas pressure in the course of cloud collapse 
and  decelerates the radial collapse significantly. 
The magnetic braking is also effective in models BL and BS.

Also in model BL the high density core changes its form from 
prolate to oblate as the central density increases, as shown 
in \fig{B1}, which is the same as \fig{A1} but for model BL.  
The change in the core shape is due to the magnetic field, 
which is amplified during the spherical collapse.
The ratio of the magnetic pressure to the gas pressure
is 0.11 at the stage of $ n _{\rm c} = 2 \times 10^6 $, while it
is only 0.05 at the initial stage.
Each panel of \fig{B1} denotes the density and velocity 
distribution at the stage of 
(a) $ n _{\rm c} \, = \, 8.2 \times 10^3 \cm $, 
(b) $ 5.6 \times 10 ^4 \cm $,
(c) $ 7.9 \times 10 ^6 \cm $, and (d) $ 6.0 \times 10 ^{10} \cm$.
At the stage of $ n _{\rm c} \, = \, 8.2 \times 10 ^3 \cm $,
the oblateness is  $ \varepsilon _{\rm ob} $ = 0.58 
in model BL [Fig. 6(a)] while $ \varepsilon _{\rm ob} $ = 0.45 
in model AL [Fig. 1(b)] .  
The core is more oblate in model BL than in models AL and AS
when compared at a given stage with the same central density
[Fig. 3(b)].
The oblateness is $ \varepsilon _{\rm ob} $ = 5.3 
at the stage of $ n _{\rm c} \, = \, 5 \times 10 ^{10} \cm $
in model BL, while $\ob =2.9$ in model AS (see Fig. 3).    
The oblateness increases slowly over the period of 
$ 3 \times 10 ^3 \cm \la n _{\rm c} \la 10 ^{10} \cm $
in model BL and is saturated around $ \ob \simeq 5 $ in the 
period of $ n _{\rm c} \ga  10 ^{10} \cm $.

As well as in models AS and AL, the axis ratio decreases
from the initial value of $ \ar = 0.2 $ to 0.015 in the
period of $ n _{\rm c} \, \la \,  10 ^4 \cm $
in model BL.  
Then it switches to growing in proportion 
to $ n_{\rm c} ^{1/6} $ in the period of $ n _{\rm c} \,
\ga \, 10 ^4 \cm $. 
The core is elliptic on the $ x - y $ plane and the
axis ratio is $ \ar $ = 0.23 at the stage of
$ n _{\rm c} \, = \, 5 \times 10 ^{10} \cm$,
as shown in the upper panel of \fig{B1} (d).
The amplitude of the non-axisymmetic perturbation
is linearly proportional to the initial amplitude.  
The axis ratio is always smaller by a factor of 20 in 
model BS than in model BL when compared at the 
stage of a given central density.
Models BS and BL have the same model parameters except 
for $ \amt $.


Since the core is appreciably oblate, the infall velocity
is higher in the vertical direction than in the radial
direction.  At the stage of 
$ n _{\rm c} = 5 \times 10 ^{10} \cm $ the radial infall 
velocity is $ v _r = - 0.46 \km$ at maximum in $ z \, = \, 0 $ plane 
while the vertical infall velocity is $v_z  = 0.7 \km $ at maximum 
on the $ z $-axis.  The radial infall velocity is smaller and the vertical infall velocity is larger than in model AS.   
This asymmetry is due to the magnetic field.  
The rotation velocity is quite small ($v_{\varphi} = 0.03 \km$) 
and the centrifugal force is negligible.
The density increase due to collapse is slower in model BL
than in model AS.  The growth of the central density is well 
approximated by 
$ \rho _{\rm c} \, = \, 2.5 / [ 4 \pi G \, (t \, - \, t _{\rm f} ) ^2 ] $.
The growth rate is 7 \% smaller than that of model AS at a given
central density.

Also in model BL, the magnetic field strengthens as the
density increases.  The ratio of the magnetic pressure
to the gas pressure 
increases very slowly in the period of 
$ n _{\rm c} \la 10 ^6 \cm $ (see Fig. 4).
It is saturated around $ (B_{\rm zc}/B_{\rm zc,0}) ^2 /(\rho_c/\rho_{c,0})  \simeq 1.3 $
in the period of $ n _{\rm c} \ga  10 ^6 \cm $.
The magnetic field decelerates the radial collapse
appreciably as shown earlier.

\fig{B2} shows the magnetic field lines at the stage of
 $ n _{\rm c} \, = \, 6 \times 10 ^{10} \cm $.  
The magnetic field lines break at the levels of $ z  \simeq 15$ AU
and 30~AU near the disk surface.  
The latter break corresponds to a fast-mode MHD shock,
which is essentially the same as the shock waves seen in
\citet{norman80}, \citet{matsu97}, and \citet{nakamura99}.  
They are squeezed and vertical to the midplane 
below the shock front while they are open above the shock front.  
The disk formation is due to the magnetic field.  

The magnetic field extracts angular momentum from the core.
As shown in \fig{bz} (b), the ratio of angular velocity 
to the magnetic field decreases by 30 \% in the period of 
$5 \times 10^2 \cm < n_c \le 5\times 10^{10}\cm$ in model BL, 
although it remains constant in model AL.
The decrease is due to the magnetic braking. 
The twisted magnetic field transfers the angular momentum
of the core outwards. 
The specific angular momentum of the core is 70 \% of the
initial value at the stage of $ n _{\rm c} = 5 \times 10 ^{10} \cm $.
The angular velocity normalized by free-fall timescale 
[$\Omega _c/(4 \pi G \rhoc)^{1/2}$] increases slightly 
from 0.01 to 0.015 in the period of  
$5 \times 10^2 < n_c \le 5 \times 10^{10} \cm$ in model BL, 
while it spins up from 0.01 to 0.06 in model AL.
We discuss this difference again in \S 4.5 in which we 
compare the increase in $ \Omega _c $ for various models.

The efficiency of the magnetic braking is qualitatively
similar in models BS and BL.

Models BL and BS are similar to model C of \citet{tomisaka95}
and model B1 of \citet{nakamura99}, although the earlier models include neither rotation 
nor non-axisymmetric perturbation.  
The rotation and non-axisymmetric perturbation have little effect on the cloud collapse in our models BL and BS.
When the initial magnetic
pressure is larger than a tenth of the gas pressure
($\alpha \geq 0.1 $), initially weak magnetic field 
is amplified during the collapse and affects the evolution
of the core.  The magnetic pressure decelerates the radial
collapse and leads to disk formation.  Also the magnetic
braking is appreciable.

\subsection{Weakly Magnetized and Rapidly Rotating Cloud}
\label{sec:modelC}

This subsection describes model CS as a typical example 
for models having small $ \alpha $ and large $ \omega $.
Model CS has parameters of $ \alpha \, = \, 0.01 $, 
$ \omega \, = \, 0.5 $, and $ \amt \, = \, 0.01 $. 
When $ \omega \, \ge \, 0.1 $, rotation affects the collapse of the 
cloud significantly.  

In model CS, a rotating disk forms at an early stage of
 low central density.
Each panel of \fig{C1} denotes the density and velocity 
distribution at the stages of 
(a) $ n _{\rm c} \, = \, 5.2 \times 10^3 \cm $, (b) $ 6.5 \times 10 ^4 \cm $,
(c) $ 5.7 \times 10 ^6 \cm $, and (d) $ 8.3 \times 10 ^{10} \cm$.
The rotating disk is clearly seen at the stage of 
$ 6.5 \times 10 ^4 \cm $.  The oblateness reaches
$ \ob = 3.0 $ at the stage of $ 4.0 \times 10 ^4 \cm $ and
is saturated around $ \ob \, \simeq \, 3.5 $ 
in the period of $ 5 \times 10 ^5 \cm \la n _{\rm c} <
5 \times 10 ^{10} \cm $ as shown in \fig{ob} (b).  

The axis ratio increases up to $ \ar \, = \, 0.2 $ 
by the stage of $ n _{\rm c} \, = \, 5 \times 10 ^{10} \cm $ 
in model CS (see Fig. 3).  
At the stage of $ n _{\rm c} \, = \, 5 \times 10 ^{10} \cm $,
the axis ratio is larger in model CS than in model AS, while it is 
the same at the initial stage.
The difference arises in the period of
$ n _{\rm c} \, \le \, 5 \times 10 ^4 \cm $.
The axis ratio remains around 0.01 
in model CS, while it decreases to $ 7 \times 10 ^{-4} $ 
in model AS.  
The axis ratio grows roughly in proportion to $ n _{\rm c} ^{1/6} $ 
in the period of $ n \, \ga \, 5 \times 10 ^4 \cm $ both in
model AS and in model BS. 
We have confirmed that the non-axisymmetric perturbation is
proportional to the initial perturbation by comparing with 
model CL of which initial parameters are the same as those of
model CS  except for $ \amt $. 
The axis ratio is 20 times larger in model CL than in
model CS in the period of $ n _{\rm c} \, \la \, 1.0 \times 10 ^9 \cm $.
The axis ratio reaches $\ar = 10.2$ and the high density
core has a bar shape  at the stage
of $ n _{\rm c} \, = \, 5 \times 10 ^{10} \cm $ in model CL.

The increase in the central density is approximated 
by $ \rho _{\rm c} \, = \, 6.2 / [ 4 \pi G \, (t \, - \, t _{\rm f}) ^2] $.
The rate of the increase is appreciably smaller than those of
models AS and BS.  It is 1.93 times smaller than that of
the spherical collapse at a given central density.
The relatively slow collapse is due to fast rotation.

\begin{figure}
\begin{center}
\includegraphics[width=140mm]{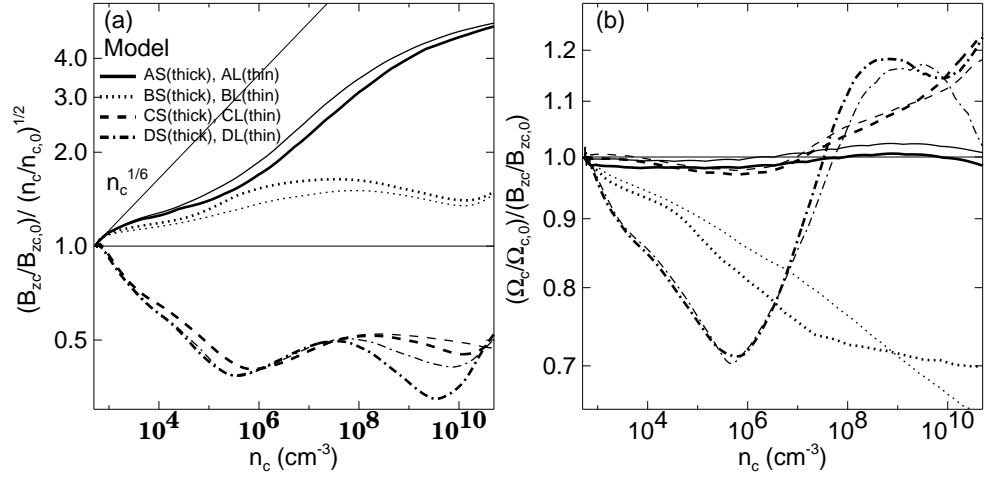}
\caption{
The square root of the ratio of the magnetic pressure to the gas pressure normalized by the initial value, $(\bzc/B_{\rm zc,0})/(n_c/n_{\rm c,0})^{1/2}$, is plotted against the central density, $n_c$, in the left panel.
The ratio of the angular velocity to the magnetic flux density normalized by the initial value, ($\Omega_c/\Omega_{c,0})/(\bzc/B_{\rm zc,0}$), is plotted against the central density in the right panel. 
The relation $(n_{\rm c}/n_{c,0})^{1/6}$ is also plotted in the left panel.
}
\label{fig:bz}
\end{center}
\end{figure}

\begin{figure}
\begin{center}
\includegraphics[width=120mm]{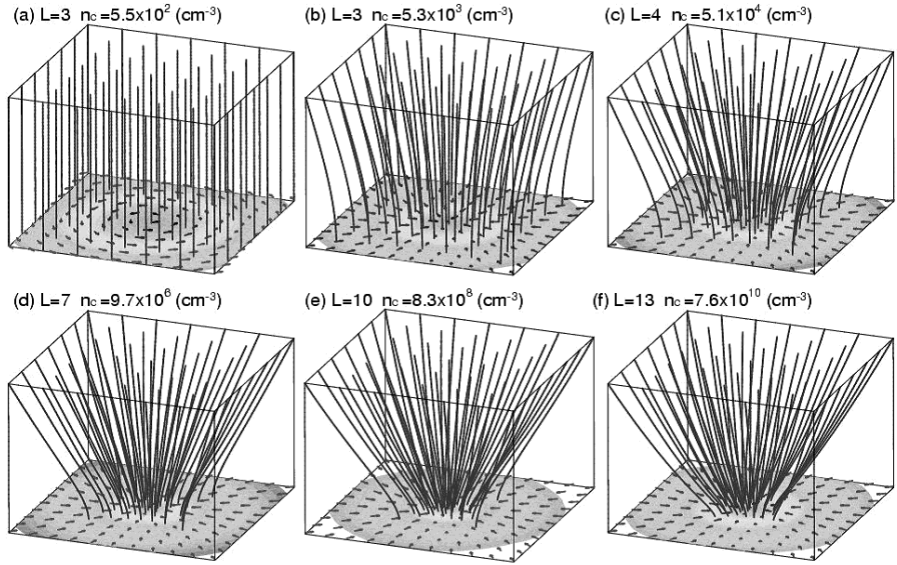}
\caption{
The magnetic field lines are shown from a bird's eye view for model AS.
Panels (a) through (f) denote the same epoch as those of \fig{A1} (a) $-$ (f), respectively.
The density (gray scale) and velocity (arrows) in $z=0$ plane are also shown on the bottom. 
The level of subgrid and central density are shown in the upper section of each panel.
Each frame denotes a cube with side lengths of
(a) $1.7 \times 10^5$ AU, (b) $1.7 \times 10^5$ AU, (c) $8.4 \times 10^4$ AU, 
(d) $1.1 \times 10^4$ AU, (e) $1.3 \times 10^3$ AU, and (f) $1.6 \times 10^2$ AU, respectively.
}
\label{fig:A2}
\end{center}
\end{figure}

\begin{figure}
\begin{center}
\includegraphics[width=170mm]{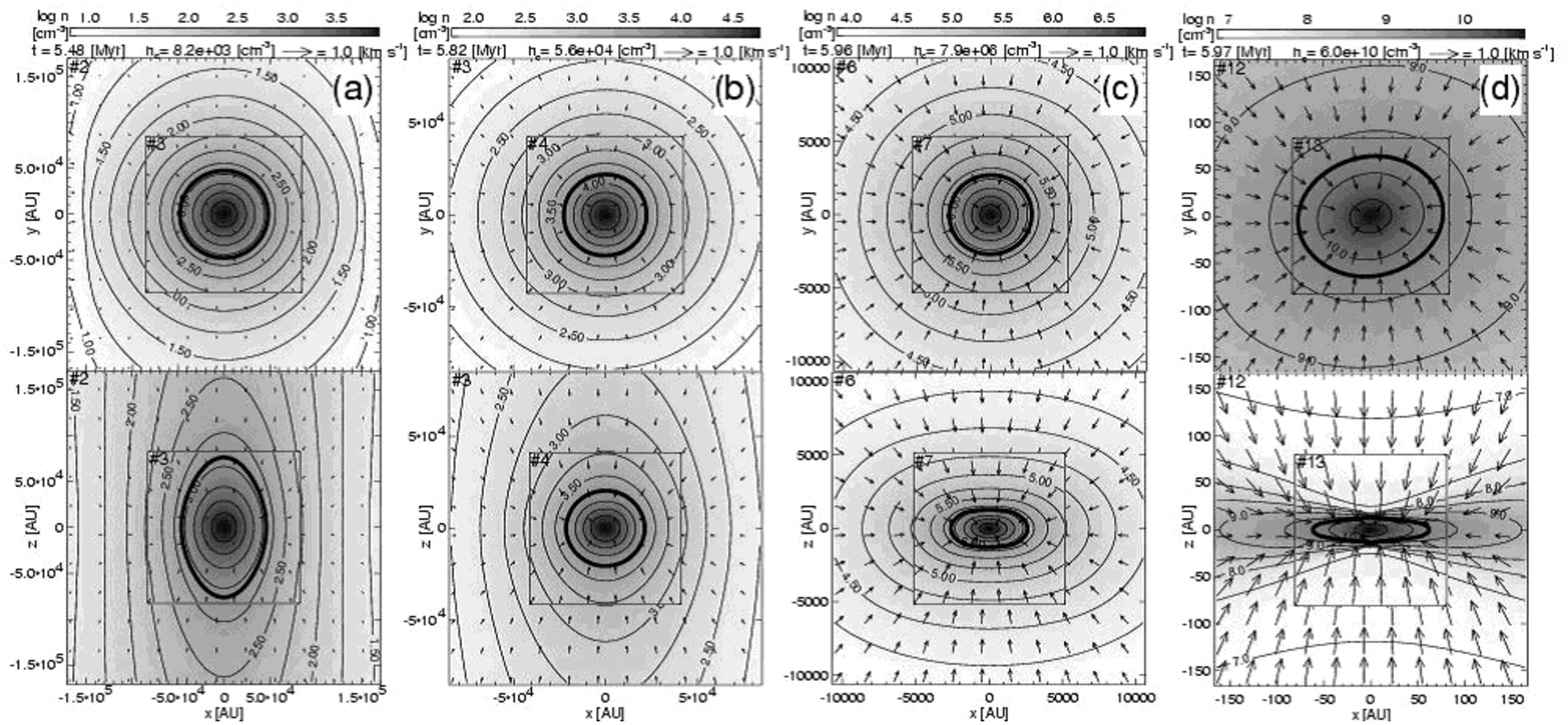}
\caption{
The density (gray scale and contours) and velocity distributions (arrows) for model BL [($\alpha$, $\omega$, $\amt$)=(0.1, 0.01, 0.2)] are shown on the $z = 0$ (upper panels) and $y = 0$ (lower panels) planes.
Panels (a) through (d) are snapshots at the stages of 
(a) $n_c = 8.2 \times 10^3 \cm$ ($l=2, 3$), \  
(b) $ 5.6 \times 10^4 \cm$ ($l=3, 4$), \  
(c) $ 7.9 \times 10^6 \cm$ ($l=6, 7$), \ and \  
(d) $ 6.0 \times 10^{10} \cm$ ($l=12, 13$), \  
The contours, arrows, and notation have the same meaning as in \fig{A1}.
}
\label{fig:B1}
\end{center}
\end{figure}

\begin{figure}
\begin{center}
\includegraphics[width=100mm]{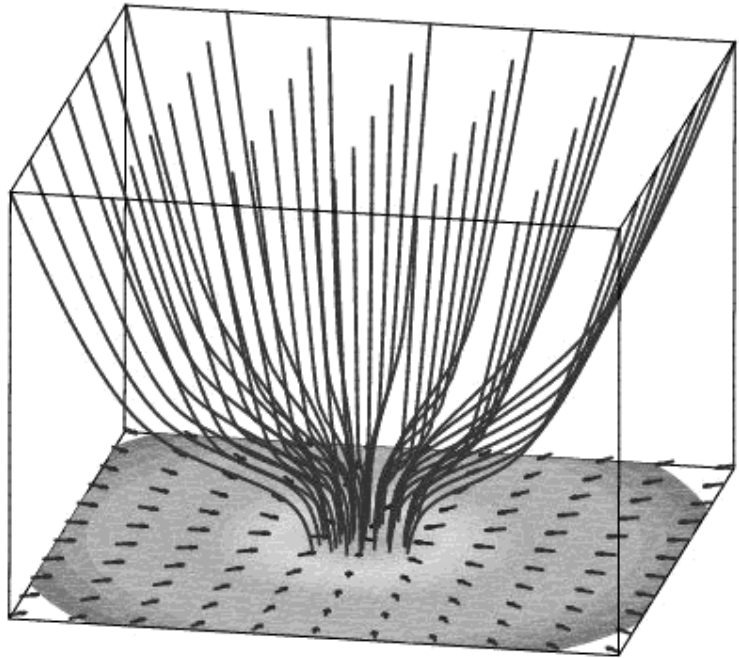}
\caption{
The magnetic field lines are shown at the same epoch of Fig.~\ref{fig:B1} (d).
The level of subgrid and central density are $l=13$ and $n_c = 6.0 \times 10^{10} \cm$, respectively.
The frame denotes a cube with a side length of 160 AU.
The gray scale and arrows have the same meaning as in \fig{A2}.
}
\label{fig:B2}
\end{center}
\end{figure}

\begin{figure}
\begin{center}
\includegraphics[width=170mm]{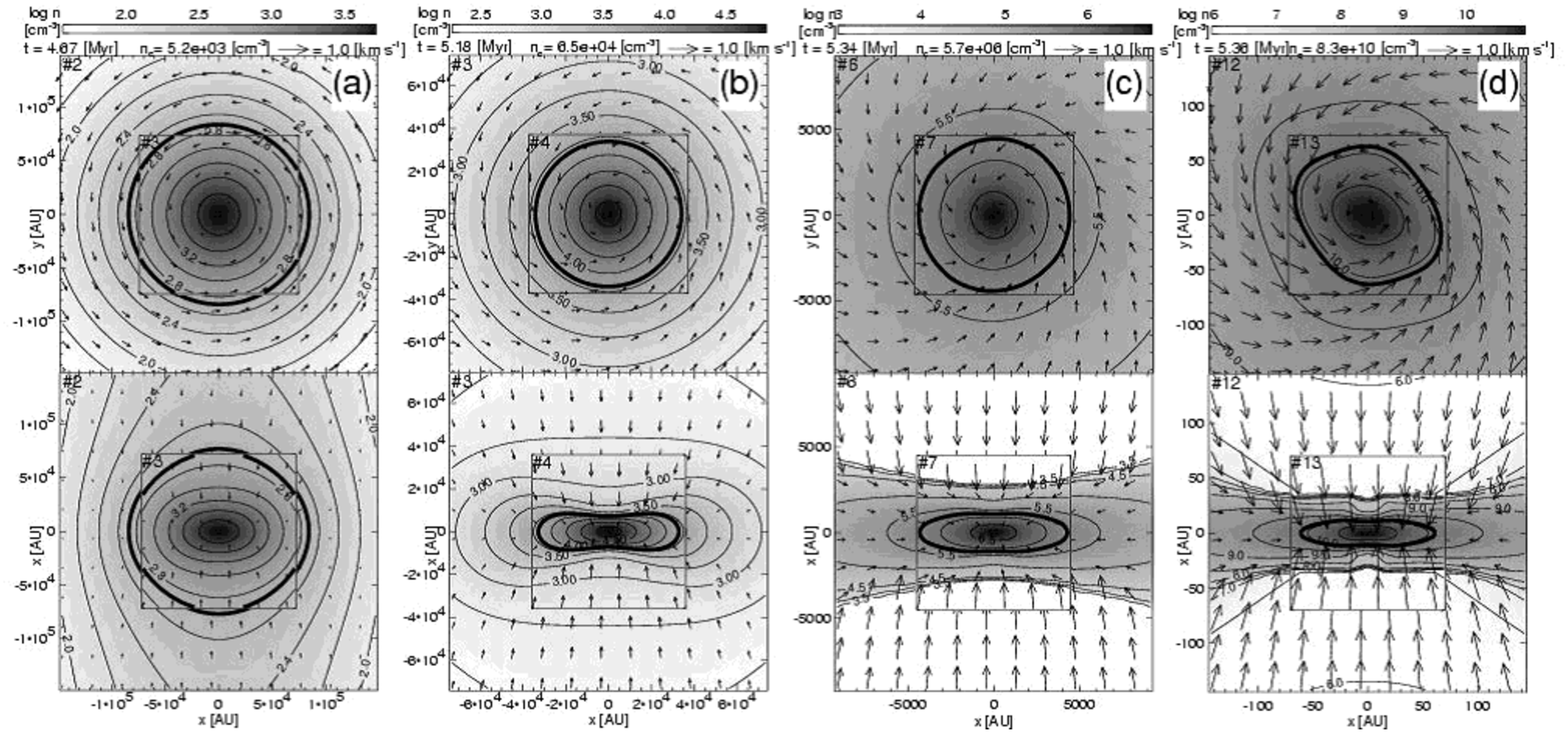}
\caption{
The density (grey scale and contours) and velocity distributions (arrows) for model CS [($\alpha$, $\omega$, $\amt$)=(0.01, 0.5, 0.01)] are shown on the $z = 0$ (upper panels) and $y = 0$ (lower panels) planes.
Panels (a) through (d) are snapshots at the stages of 
(a) $n_c = 5.2 \times 10^3 \cm$ ($l=2, 3$), \  
(b) $6.5 \times 10^4 \cm$ ($l=3, 4$), \  
(c) $5.7 \times 10^6 \cm$ ($l=6, 7$), \ and \  
(d) $8.3 \times 10^{10} \cm$ ($l=12, 13$), \  
The contours, arrows, and notation have the same meaning as in \fig{A1}.
}
\label{fig:C1}
\end{center}
\end{figure}

In the period of $ n _{\rm c} \la 5 \times 10 ^5 \cm $ the
cloud collapses mainly in the vertical direction along 
the magnetic field. 
Accordingly the magnetic field increases a little and
the ratio of the magnetic pressure to the gas pressure
decreases in this period (see \fig{bz}).
Note that the square root of the ratio of the magnetic 
pressure to the gas pressure decreases in proportion to 
$ \rhoc ^{-1/2} $ when the collapse is purely
vertical along the magnetic field.
Also, the angular velocity increases a little and
decreases in proportion to $ \rhoc ^{-1/2} $
when measured in the freefall timescale.

In the period of $5\times 10^2 \cm < n_c \la 5\times 10^{10} \cm$,
the magnetic field ($B_{\rm zc}$) strengthens and the angular velocity  
of the core ($\Omega_{\rm c}$) continue to increases.  
However, the ratio of the magnetic
pressure to the gas pressure remains nearly constant 
around $  (B_{\rm zc}/B_{\rm zc,0}) ^2 /(\rho_c/\rho_{c,0}) 
 = 0.5 $ for $5\times 10^5 \cm < n_c \la 5\times 10^{10} \cm$ as shown in \fig{bz}.
The angular velocity measured in the freefall timescale is
also nearly constant around $ \Omega _c (4 \pi G \rhoc) ^{-1/2} =0.2 $.
In other words, both the magnetic field and angular velocity 
increase in proportion to $ \rhoc ^{1/2} $.  
These dependences of $ B _{zc} $ and $ \Omega _{\rm c} $
on $ \rhoc $ indicate that the core collapses in the radial
direction while maintaining a disk shape.
They are the same as those in the similarity
solution for a self-gravitationally collapsing gas disk
(Tomisaka 1995, 2002; Nakamura et al. 1995; Matsumoto \etal 1997; Saigo \& Hanawa 1998).

\begin{figure}
\begin{center}
\includegraphics[width=170mm]{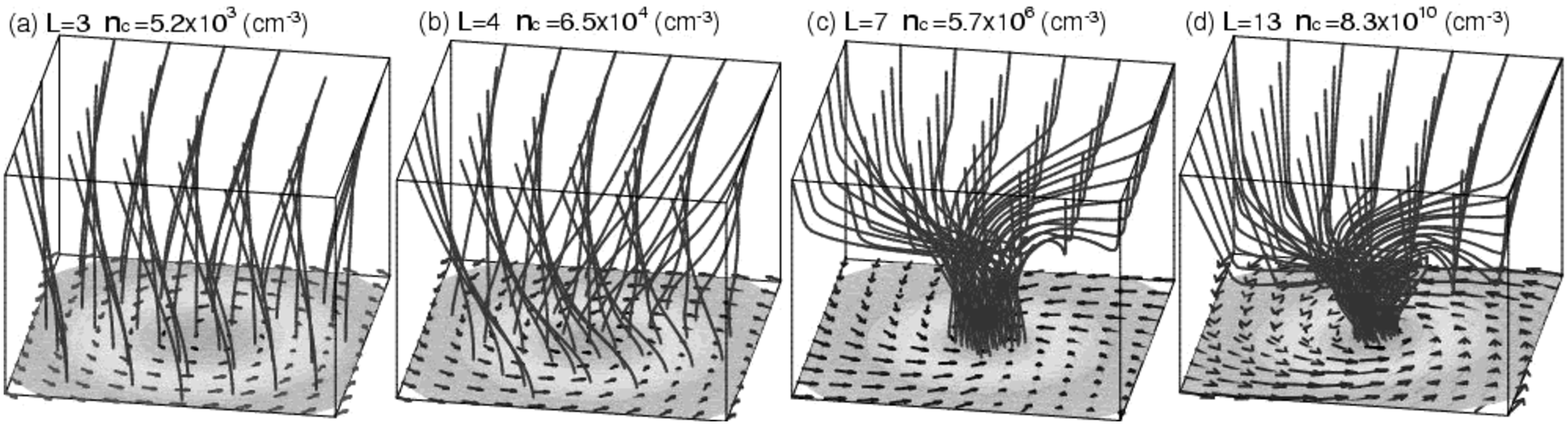}
\caption{
The magnetic field lines at the same epoch of Fig.~\ref{fig:C1}.
Each frame denotes a cube with side lengths of  
(a) $1.7 \times 10^5$ AU, (b) $8.4\times 10^4$ AU, 
(c) $1.1\times 10^4$ AU, and (d) $1.6\times 10^2$ AU, respectively. 
The gray scale and arrows have the same meaning as in \fig{A2}.
}
\label{fig:C2}
\end{center}
\end{figure}

The ratio of the angular velocity to the magnetic field is constant 
in the period of $ n _{\rm c}  < 10^7 \cm$.
It increases up to 1.2 by the stage of $n_c = 5 \times 10^{10} \cm$.
This increase is due to the torsional Alfv\'en wave.
The magnetic braking is not significant in model CS.
The ratio of the angular velocity to the magnetic field is 
constant during the collapse as shown in \fig{bz} (b). 
This confirms that the specific angular momentum is
conserved.   The magnetic field is twisted by fast rotation 
as shown in Fig.~\ref{fig:C2}.    
It is also bent at the shock front as well as in model BS. 
The twisted magnetic field is too weak to have any 
appreciable dynamical effects.

The infall velocity is higher vertically than radially.  
The maximum infall velocity is $ v _{r,{\rm max}} \, = \, - 0.37 \km $ radially and $ v _{z,{\rm max}} \, = \, \pm 0.59 \km $ at the stage
of $ n _c = 5 \times 10 ^{10} \cm $.  The maximum rotation
velocity is $ v _{\varphi} = 0.36 \km $ and exceeds the sound
speed at the same stage.  Thus both the infall and rotation
are supersonic.  
This dynamically infalling gas disk is
similar to infalling envelopes observed in HL Tau \citep{hayashi93} and L1551 IRS5 \citep{ohashi96,saito96} in the sense that the radial infall velocity is comparable
with the rotation velocity.
The vertical inflow along the $ z $-axis forms shock waves  
twice, once at the stage of $ n _{\rm c} \, = \, 5.7 \times 10^6 \cm $ 
[see \fig{C1} (c)] and at that of 
$ n _{\rm c} \, = \, 8.3 \times 10 ^{10} \cm $ 
[see \fig{C1} (d)].  The former forms  at 
$ z \, = \, \pm 4 \times 10 ^3 $~AU, and 
the latter at $ z \, = \pm \, 40 $~AU.
These shock waves are essentially the same as those seen
in Matsumoto et al. (1997).
The oblateness has a temporal maximum value at the stages of 
the shock formation.

\subsection{Strongly Magnetized and Rapidly Rotating Cloud}
\label{sec:modelD}

This subsection describes model DL as a typical example of models 
having large $\alpha$ and large $\omega$.
Model DL has parameters $\alpha=1.0$, $\omega=0.5$ and $\amt=0.2$.
When $\alpha \ge 0.1$ and $ \omega \, \ge \, 0.1 $, 
both magnetic field and rotation affect the collapse of 
the cloud significantly.
The magnetic braking is also effective.

\begin{figure}
\begin{center}
\includegraphics[width=190mm]{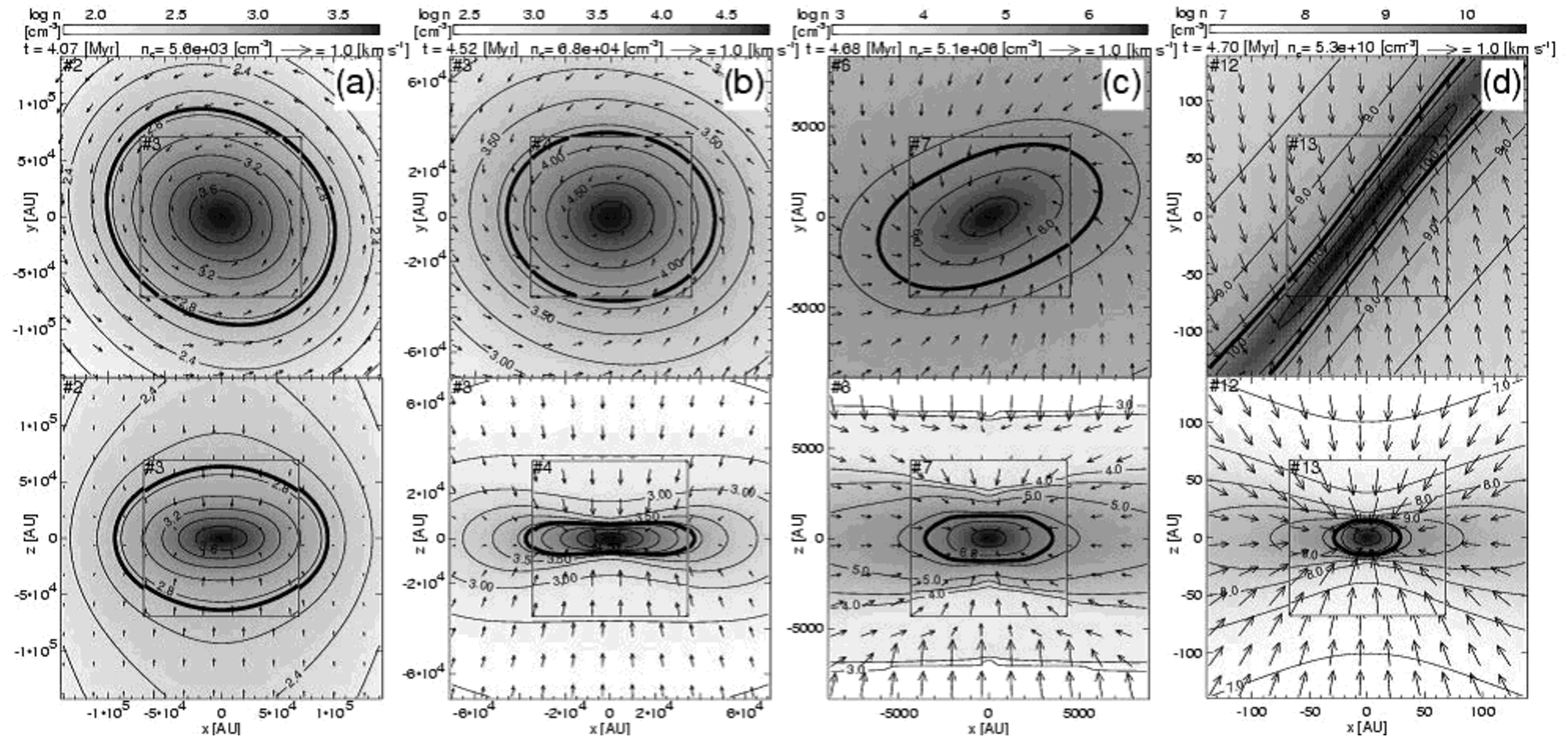}
\caption{
The density (grey scale and contours) and velocity distributions (arrows) for model DL [($\alpha$, $\omega$, $\amt$)=(1, 0.5, 0.2)] are shown on the $z = 0$ (upper panels) and $y = 0$ (lower panels) planes.
Panels (a) through (d) are snapshots at the stages of 
(a) $n_c =5.6 \times 10^3 \cm$ ($l=2, 3$), \  
(b) $6.8 \times 10^4 \cm$ ($l=3, 4$), \  
(c) $5.1 \times 10^6 \cm$ ($l=6, 7$), \ and \  
(d) $5.3 \times 10^{10} \cm$ ($l=12, 13$), \  
The contours, arrows, and notation have the same meaning as in \fig{A1}.
}
\label{fig:D1}
\end{center}
\end{figure}

\fig{D1} shows formation of a magnetized rotating disk that deforms to an elongated high density bar in model DL.  
Each panel denotes the density and velocity distribution at the stage of 
(a) $ n _{\rm c} \, = \, 5.6 \times 10^3 \cm $, 
(b) $6.8 \times 10 ^4 \cm $,
(c) $ 5.1 \times 10 ^6 \cm $, and (d) $ 5.3 \times 10 ^{10} \cm$.
The high density gas has an oblateness of $ \ob = 4.2 $ at the stage
of $ n_c = 6.8 \times 10^4 \cm $.
The oblateness reaches its maximum at $n_{\rm c} \simeq 4\times 10^5 \cm$ and
oscillates around $\ob\simeq 5$ in the period of 
$ 4 \times 10 ^5 \cm \, \la \, n_{\rm c} \, \le \, 5\times 10^{10} \cm$ 
(see \fig{ob}).  As a result of the strong magnetic field and fast 
rotation, the disk forms at an earlier stage in model DL than in
the other models shown in the previous subsections.

The increase in the central density is approximated by 
$n_{\rm c} = 4.9 / \vert 4 \pi\, G \, (t-t_{\rm f})^2 \vert$.
This is slower than in models AS and BS, although faster than in model CS.
The density increase is faster in a model having a larger $ \alpha $
for a given $ \omega \, \ge \, 0.1 $.  This is because a stronger
magnetic field brakes the rotating core more effectively and
the centrifugal force is reduced more. 
Remember that the density increase is slower in a model having 
a larger $ \alpha $, when $ \omega \ll 0.1 $.
When the angular momentum of the cloud is very small,
the centrifugal force is negligible and its reduction
due to the magnetic braking is unimportant.  A stronger 
magnetic field decelerates the collapse through higher
magnetic pressure and tension.  The magnetic field 
plays two roles: acceleration of the collapse through 
magnetic braking, and  deceleration of the collapse through
magnetic pressure and tension.  The former dominates for 
$ \omega \ge 0.1 $ while the latter dominates for $ \omega < 0.1 $.

As shown in \fig{D1} (d) upper panel, the disk is elongated into
a bar of $ \ar = 15 $ at the stage of $ 5.3 \times 10 ^{10} \cm $.
As well as in model CL, the axis ratio remains nearly constant at
the beginning and increases in proportion to $ n _{\rm c} ^{1/6} $
from an early stage of $ n _{\rm c} < 5 \times 10 ^3 \cm $ in
model DL (see \fig{ob}).  An elongated bar forms in the models in which
the non-axisymmetric perturbation is relatively large ($ \amt > 0.2$)
and does not diminish in the early phase. 
Remember that the axis ratio decreases in the period of
$ n _{\rm c} \la 1.0 \times 10 ^4 \cm $ in models AL and BL.
The axis ratio increases in proportion to $ n _{\rm c} ^{1/6} $
in all models while the core collapses dynamically.
The final axis ratio depends on the initial value and the
amount of damping in the early phase.
The initial damping is smaller when either the initial
magnetic field or rotation is larger.

A similar bar structure is also seen in \citet{durisen86} and 
\citet{bate98}.  The bar structure develops as a result of the
bar mode instability, when the ratio of 
rotational to gravitational energy ($\beta$) of the core exceeds $\beta = 0.274$. 
This $\beta$ is related to $\omega$ by $\beta = \omega^2$, when
the cloud is spherical and has constant density and angular
velocity.  Thus, the criterion for the `bar mode instability', $\beta > 0.274$, 
corresponds to $\omega >0.523$.  The value of $\omega$ continues to decrease until it converges to $\omega \simeq 0.2$ as denoted in the following subsection.
Therefore, the condition  $\omega > 0.523$ is never fulfilled, and the
bar should form in model D by another mechanism.
While the bar mode instability of \citet{hanawa99} works
only in a dynamically collapsing cloud,  the bar mode instability
of \citet{durisen86} and \citet{bate98} does in a cloud in hydrostatic
equilibrium.

There is another evidence that the bar is formed not by fast
rotation in our models.  The bar forms also in the model of
($\alpha$, $\omega$, $\amt$) = (3,~0,~0.2), which is listed
as model 56 in Table 2 of Paper II.  The bar formation can not
be due to rotation since the cloud does not rotate in this model.
(See Table 2 of Paper II for the list of the models in which
the bar forms at the end of the isothermal phase.)


As well as in models CS and CL, the vertical infall dominates
over the radial infall in the period of 
$ n _{\rm c} \la 7 \times 10 ^4 \cm $ in  models DS and DL.
The ratio of the magnetic pressure to the gas pressure normalized by its initial value
decreases  to 0.5 in the period as shown in \fig{bz}.  Then it
oscillates around 0.5 in the period of 
$10^5 \la n_c \la 5 \times 10^{10} \cm $.  
The epoch of disk formation coincides with that at which the
ratio of the magnetic pressure to the gas pressure reaches its first local
minimum value.

The vertical inflow also forms shock waves twice in model DL.
\fig{D1}(c) shows the outer shock located at $z = \pm 7000$ AU.
The flow is nearly vertical above the front while it is horizontal 
below.  
The epoch of shock formation coincides with that of the temporarily maximum oblateness as in model CL.

The magnetic braking slows the spin of the collapsing disk in model DL.  
The initial central angular velocity is 
$\Omega_c/(4 \pi G \rhoc)^{1/2} = 0.5 $ in both models CL and  DL.  
The central angular velocity decreases to 
$\Omega_c/(4 \pi G \rhoc)^{1/2} = 0.21 $  
by the stage of $ n _{\rm c} \, = \, 5 \times 10 ^{4} \cm $ 
in model DL, while it decrease to 0.24 in model CL.  
The ratio of the angular velocity to the magnetic field
($ \Omega _c/ B _{zc} $) normalized by the initial value ($\Omega_{\rm c,0}/B_{\rm zc,0}$) decreases to 70 \% of the
initial value in model DL (see \fig{bz}).
The magnetic braking is effective in the period of
$ n _{\rm c} \leq  7 \times 10^5 \cm $ as in model BL.
The ratio of the angular velocity to the magnetic
field increases in the period of $ 7 \times 10^5 \cm < n _{\rm c} < 5 \times 10^8 \cm$.
This spin is due to the magnetic torque.  The twist of
the magnetic field is bounded by the shock front and the
torsional Alfv\'en wave is reflected there.  Thus
the angular momentum is not released from the core
in model DL.

\begin{figure}
\begin{center}
\includegraphics[width=170mm]{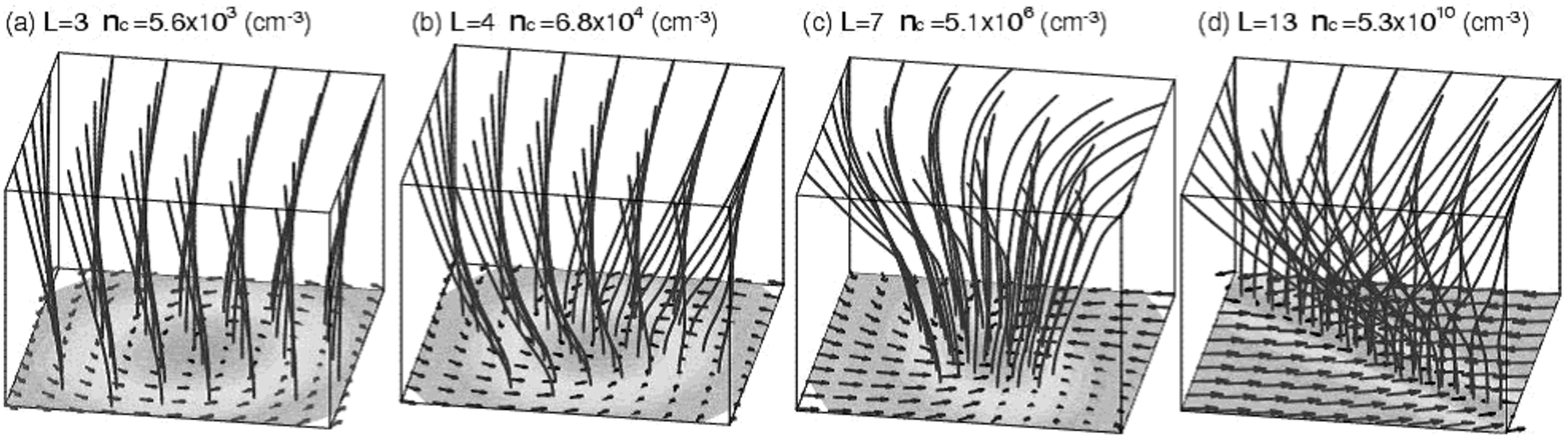}
\caption{
The magnetic field lines at the same epoch of Fig.~\ref{fig:D1}.
Each frame denotes a cube with side lengths of 
(a) $1.7 \times 10^5$ AU, (b) $8.4\times 10^4$ AU, 
(c) $1.1\times 10^4$ AU, and (d) $1.6\times 10^2$ AU, respectively.
The gray scale and arrows have the same meaning as in \fig{A2}.
}
\label{fig:D2}
\end{center}
\end{figure}

The ratio of the magnetic pressure to gas pressure decreases from 
0.5 to 0.1 in the period of $5\times 10^2 \cm \le n_{\rm c} 
\la 7\times 10^5 \cm$, and remains around 0.1 in the period of 
$n_{\rm c} \ga 7 \times 10^5 \cm $.  Thus, importance of
the magnetic force relative to the centrifugal force 
decreases in models DS and DL. 
Fig.~\ref{fig:D2} illustrates the magnetic field for the stages 
shown in Fig.~\ref{fig:D1}.  
The magnetic field lines are twisted but less pinched than in model CS.
They are twisted at a higher $ z $ in model DL than in model CS.
As shown in \fig{D2} (d), the magnetic field is squeezed to stem 
vertically from the bar and the magnetic flux density is
large in the bar.
In \fig{D2} (d), magnetic field lines are bent at $z \simeq 40$AU, which corresponds to the shock front.
Inside the shock front, the magnetic field lines are ran vertically and hardly twisted, while twisted moderately outside of the shock front.

Models DL and DS have the same initial model parameters except
for $ \amt $, which is smaller by a factor of 20 in model DS.
As a result, only the axis ratio differs appreciably
between models DL and DS.  A high density disk is seen at
the stage of $ n _{\rm c} = 5 \times 10 ^{10} \cm $ in model
DS while the elongated bar is seen in model DL.
The axis ratio is smaller by a factor 20 in model DL 
throughout the evolution.

\subsection{Magnetic Flux - Spin Relation}
\label{sec:UL}


The filamentary cloud fragments to form a high density
core of $ n _{\rm c} > 5 \times 10 ^{10} $ in all the
models computed.  The formation of the core is dynamical
and the central density increases in proportion to
the inverse square of the time.  As the central density increases,
the magnetic field increases roughly in proportion to a power 
of $ \rhoc $.  The power index, however, differs and
depends on the geometry of the collapse as shown in the 
previous subsections.  Also the angular velocity increases
in proportion to a power of $ \rho $ and the power index
depends on the geometry of the collapse and the strength
of the magnetic field.  
To summarize the increase in $ \bzc $ and $ \Omega _c $
we have plotted the evolutionary locus of the core in
Fig.~12.   The abscissa denotes the
square root of the ratio of the magnetic pressure to the 
gas pressure, $ \bzc (8 \pi c _{\rm s} ^2 \rhoc) ^{-1/2} $,
in the logarithmic scale.
The ordinate denotes the angular velocity normalized by
the freefall timescale, $ \Omega _c (4 \pi G \rhoc)^{-1/2} $, on the logarithmic
scale.  Each curve denotes the evolutionary locus for 
a model.  The asterisks denote the initial stages.
The circles, squares and triangles denote the stages
of $ n _{\rm c} = 5 \times 10 ^4 \cm$, $ 5 \times 10 ^6 \cm$, 
and $ 5 \times 10 ^8 \cm $, respectively.  
Models without magnetic field are shown 
inside the upper left box.
Models without rotation are shown inside the lower right box.

The evolutionary loci are systematically ordered in
Fig.~12.  They are aligned to evolve toward the upper right 
in the lower left region.
Models AL and AS belong to this region of weak
magnetic field and slow rotation.   
On the other hand, the loci are
aligned to evolve toward the lower left in the upper
region (fast rotation) and in the right region 
(strong magnetic field).  Models CL, CS, DL, and DS belong
to these regions.  Models BL and BS appear in the middle
of the panel.  Their loci are nearly horizontal and the
angular velocity measured on the freefall timescale
does not increase as a result of the magnetic braking.

We can deduce several rules for the collapse of a
magnetized rotating gas cloud from Fig.~12.
First all the loci seem to converge on the curve,
\begin{equation}
\dfrac{\Omega_c^2}{(0.2)^2 \; 4 \pi G \rhoc} +
 \dfrac{B_{\rm zc}^2}{(0.36)^2 \; 8 \pi c_s^2 \rho_c} =1.
\label{eq:UL}
\end{equation}
Equation~(\ref{eq:UL}) is denoted by the thick solid curve in
Fig.~12.  We call this curve the magnetic flux - spin
relation or $ B - \Omega $ relation in
the following.  
The first term of equation~(\ref{eq:UL}) is proportional
to the square of the angular velocity normalized by the
freefall timescale and accordingly is proportional to
the ratio of the centrifugal force to the gravity. 
The second term is proportional to the ratio of the
magnetic pressure to the gas pressure.  
The numerators are proportional to the anisotropic
forces which suppress only the radial infall.  
Convergence to equation~(\ref{eq:UL}) indicates that
the sum of the centrifugal and magnetic forces are
regulated to be at a certain value.  This rule 
involves the rules found by Matsumoto et al. (1997)
and Nakamura et al. (1999) as a special case.
The former showed that the ratio of the specific
angular momentum to the core mass 
converges to a half of the critical value
$ j = 0.5 (2 \pi G M/ c _{\rm s} ) $
for models having no magnetic field.
The latter showed that the ratio of the magnetic
field to the surface density tends to be a half
of the critical one, i.e., $ \bzc = 0.5 
(2 \pi G) ^{1/2} \Sigma $, for collapse of a 
non-rotating cloud.
See the models shown inside the upper left box and those
inside the lower right box to confirm that they also
converge to equation~(\ref{eq:UL}).  

The magnetic flux - spin relation is related to formation of the shock
waves.
The first shock wave forms exactly when the 
evolutionary locus reaches the $ B - \Omega $ relation.
After the shock formation, the evolutionary locus 
leaves it temporarily and reaches it again at the formation of
the second shock wave.
The shock strength is also related to the distance between 
the initial stage and the $ B - \Omega $ relation on the
$ B - \Omega $ diagram.  
The shock wave is stronger in a model starting
from a more distant place from equation~(\ref{eq:UL}) on
the diagram.  
No shock wave forms in a model 
of which the initial stage is close to equation~(\ref{eq:UL})
on the diagram (see, e.g., the model of $\alpha=0.1$ and 
$\omega=0.1$ shown in Fig. 12).

Second, the magnetic braking is appreciable only 
in models having $ \alpha \simeq 0.1 $.
The effect of the magnetic braking is evaluated
from the slope on the diagram,  
$ d\log \Omega _{\rm c} /d \log \bzc $.
When the specific angular momentum is conserved,
the slope is $ d\log \Omega _{\rm c} / d \log \bzc = 1 $
as discussed in subsection 4.3.  
The slope differs appreciably from unity near the lower part of 
the $ B - \Omega $ relation in Fig.~12.
It is appreciably smaller than unity on the left-hand side
of the $ B - \Omega $ relation, while it
is appreciably larger on the right-hand side.
When the initial magnetic field is very weak, its magnetic torque
is negligible.  
When the initial magnetic field is strong, the vertical collapse 
    dominates.  The magnetic braking reduces the specific angular momentum
    by $ 30-40 \% $ by the stage of $ n _c \, = \, 10 ^6 \cm $.  However,
    it does not operate effectively beyond the stage.
We will discuss the implication of equation~(\ref{eq:UL}) in
the next section.

\section{Discussion}

As shown in the previous section, the magnetic flux density 
and angular velocity converge on equation~(\ref{eq:UL}) 
in Fig.~12, the $ B - \Omega $ diagram. 
Thus, we can evaluate the magnetic flux density and angular 
velocity of the first core to be
\begin{equation}
\left( \dfrac{\Omega_c}{2.57 \times 10^{-3} \, {\rm yr} ^{-1}} \right) ^2 +
\left(\dfrac{B_{\rm zc}}{1.50 \times 10^4 \, \mu{\rm G}} \right)^2 =1 ,
\label{eq:UL2}
\end{equation}
by substituting $ \rhoc \, = \, 1.92 \times 10^{-13}$ g$\cm$
(equivalent to $ n _{\rm c} = 5 \times 10 ^{10} \cm $) and
$ c _{\rm s} \, = \, 0.19~{\rm km}~{\rm s}^{-1} $, into
equation~(\ref{eq:UL}).  equation~(\ref{eq:UL2}) implies that the
first core has either the \lq standard' magnetic flux density 
(15~mG) or the \lq standard' angular velocity ($2.57\times 10^{-3}$~yr$^{-1}$),
unless the initial magnetic field is very weak and
the rotation is very slow.
When both the magnetic flux density and angular velocity
are negligible, the cloud collapses almost spherically
and hence both $ \bzc $ and $ \Omega _{\rm c} $ increase
in proportion to $ n _{\rm c} ^{2/3} $.  Thus, either 
the magnetic flux density or the angular velocity reach
the standard value at $ n _{\rm c} = 5 \times 10 ^{10} \cm $
when either $ B  \ge 1.50~\mu{\rm G} $ or 
$ \Omega _{\rm c} \ge 2.57\times 10^{-7}~{\rm yr}^{-1} $ at 
the stage of $ n _{\rm c} \, = \, 5 \times 10 ^{4} \cm$.  
The latter condition is equivalent to
$ \vert \mbox{\boldmath{$\nabla$}} \times \mbox{\boldmath$v$} \vert
\,\ge \, 2.49 \times 10^3~{\rm km}~{\rm s}^{-1}~{\rm pc}^{-1} $.

The standard magnetic flux density is approximately 
a half of the critical one, as mentioned in subsection 4.4.
The latter is evaluated to be
\begin{eqnarray}
B _{\rm cr} & = & 2 \pi \sqrt{G} \Sigma \\
& = & \sqrt{8 \pi \rho c _{\rm s} ^2}
\end{eqnarray}
at the limit of the geometrically thin self-gravitationally
bound gas disk \citep{nakano78}.  
Also, the standard angular velocity is approximately
a half of the critical one.   The critical angular
velocity is defined so that the centrifugal force
balances with the gravity.  Then it is evaluated to be
\begin{equation}
\Omega _{\rm cr} \, = \, 
\sqrt{ \dfrac{4 \pi G \rho}{3}} 
\end{equation}
for a uniform gas sphere.
Thus, equation~(\ref{eq:UL}) implies that either the magnetic flux
density or the angular velocity is regulated to be a half
of the critical value.

Equation~(\ref{eq:UL}) also predicts anti-correlation between the 
magnetic flux density and angular velocity of the
first core.  In other words, only one of the magnetic flux density 
or the angular velocity is close to the standard value.
Then we can make a new index, the ratio of angular velocity to
the magnetic flux density, for identifying whether the magnetic
field dominates over rotation during the cloud collapse.
If it is larger than the ratio of the standard values,
\begin{eqnarray}
\dfrac{\Omega _{\rm st}}{B _{\rm st}} & = & 
0.39 \, \sqrt{G}{ c _{\rm s}}^{-1} \\
& = & 1.69\times 10^{-7} \, \left( \frac{c _{\rm s}}{0.19~{\rm km~s}^{-1}} 
\right) ^{-1} \, {\rm yr}^{-1} \, \mu {\rm G}^{-1} ,
\label{eq:UL3}
\end{eqnarray}
the centrifugal force dominates over the magnetic force.
Otherwise the magnetic force dominates over the centrifugal
force.  This analysis suggests that there exist two types of first core:
magnetic first core and spinning first core. 
We discuss the difference between them in Paper II.

We shall apply the above discussion to L1544, the 
prestellar core, of which rotation and magnetic 
field have been measured.
The rotation velocity is evaluated to be 
$ 0.09~{\rm km~s}^{-1} $ at $ r $~=~15000~AU by \citet{ohashi99}
and $ 0.14~{\rm km~s}^{-1} $ at $ r $~=~7000~AU by \citet{williams99}.
These velocity gradients correspond to $ 1.26 \times 10 ^{-6} $~yr$^{-1}$
and $ 4.21 \times 10 ^{-6} $~yr$^{-1}$.
On the other hand, the line-of-sight magnetic field is evaluated to be 
$ + 11 \pm 2 \, \mu{\rm G}$ by \citet{crutcher00}.
Combining these values, we obtain $ \Omega / B $ =
$ 1.1 \times 10 ^{-7} $~yr$^{-1}$~$\mu{\rm G}^{-1}$
and 
$ 3.8 \times 10 ^{-7} $~yr$^{-1}$~$\mu{\rm G}^{-1}$.
If we take account of uncertainty of the observed values, the magnetic force dominates over
the centrifugal force only marginally.  

It should be noted that \citet{crutcher04}
derived a much stronger magnetic field ($ \approx 140~\mu$G)
for L1544 from linear polarization of the dust emission.
They derived the value under the assumption that 
the randomness of the magnetic field can be ascribed to turbulent
motion.  If the magnetic field is as strong as 140~$\mu$G 
at the distance of 10000~AU, the magnetic force should dominate 
over the centrifugal force.  However, their method gives a magnetic field
an order of magnitude stronger compared with
the values derived by the Zeeman effect.  Possible systematic
errors should be examined.

Next, we discuss the speed of dynamical collapse in the
molecular cloud core.  \citet{aikawa01} discussed the 
possibility of deriving the collapse speed from the chemical
abundance in the prestellar core L1544.  
They computed chemical evolution in a molecular cloud
core, assuming that the density evolution is the same as
that of the Larson-Penston similarity solution, or by a
factor $ f $ slower.  
They concluded that the observed chemical anomaly in L1544
is consistent with the model based on the Larson-Penston
similarity solution from comparison with the slow collapse 
models of $ f $ = 3 and 10.  
The model of $ f $ = 3 is supposed to mimic a molecular 
cloud core of which collapse is slowed down owing to 
rotation, magnetic field, or turbulence.  Our simulation
has shown that the slowing by magnetic field and rotation
is appreciably smaller.  The slow-down factor is evaluated
to be $ f $ = 1.22 for model BS and 1.93 for model CL.
The small slow-down factor makes the chemical diagnosis
harder.

Finally, we discuss the effect of the ambipolar diffusion.
The evolution of magnetically subcritical cloud including the ambipolar diffusion has been investigated by Basu \& Mouschovious (1994, 1995a, 1995b) under the disk approximation.
They showed that the magnetically supercritical core is formed in the subcritical cloud for ambipolar diffusion after 10-20 freefall time passed. 
Once the supercritical core is formed, the magnetic field is hardly extracted from the core, because the ambipolar diffusion is much slower than the freefall \citep{basu94}.
Thus, our ideal MHD approximation is valid since our model cloud is 
supercritical from the initial stage (see Table 2).  
The ambipolar diffusion may have
an important role in the initially subcritical cloud and after the
formation of the dense core ($ n _c > 10 ^{11} \cm $).

\section*{Acknowledgments}
We have greatly benefited from discussion with Prof.~ M. Y.~ Fujimoto,  
Prof.~ A.~ Habe and Dr.~K. Saigo.
Numerical calculations were carried out with a Fujitsu VPP5000 
at the Astronomical Data Analysis Center, 
the National Astronomical Observatory of Japan.
This work was supported partially by 
the Grants-in-Aid from MEXT (15340062, 14540233 [KT], 
16740115 [TM]).

\clearpage



\begin{table}
\begin{center}
\caption{Model Parameters}
\label{table:para}
\begin{tabular}{cl}
\hline
parameter & values \\
\hline 
$\alpha$  & 0,\, 10$^{-3}$,\, 5$\times$10$^{-3}$,\,  
          0.01,\,  0.05,\, 0.1,\, 0.5,\, 1,\, 2,\, 3 \\
$\omega$  & 0,\, 0.01,\, 0.02,\, 0.03,\, 0.04,\, 0.05,\, 
0.1,\, 0.2,\, 0.3,\, 0.4,\, 0.5,\, 0.6\\
$\amt$ & 10$^{-3}$,\, 0.01,\, 0.1,\, 0.2,\, 0.3\\
$n_{\rm c,0}$ & $5 \times 10^2 \cm$,\, $5 \times 10^4 \cm$,\, $5 \times 10^6 \cm$ \\
\hline
\end{tabular}
\end{center}
\end{table}

\begin{table}
\setlength{\tabcolsep}{4pt}
\caption{Parameters and Initial Conditions for Typical Models} 
\label{table:init}
\begin{center}
\begin{tabular}{ccccccccccccccccc}
\hline
 \multicolumn{2}{c}{Model}  
& $\alpha$  & $\omega$   & $\amz$  
& $\amt$ & $n_{\rm c,0}$ 
& $B_{\rm zc,0}$ {\scriptsize ($\mu$G) }
& $\Omega_{\rm c,0}$ ({\scriptsize $10^8$ yr$^{-1}$}) & $M$ {\scriptsize($\msun$)}
& L {\scriptsize ($10^5$ AU)}  & $t_{\rm f}$\ {\scriptsize (10$^6$ yr)}
& $M/M_{\rm B,cri}$ \\
\hline 
&AS& 0.01& 0.01& 0.1& 0.01&  $ 5 \times10^2$&0.295& 1.26& 12.2& 6.92 & 5.96 & 13.2 \\
\raisebox{0.5\normalbaselineskip}[0pt][0pt]{A}
&AL& 0.01& 0.01& 0.1&  0.2& $ 5 \times10^2$ &0.295& 1.26& 12.2& 6.92 & 5.99 & 13.2 \\
&BS&  0.1& 0.01& 0.1& 0.01& $ 5 \times10^2$ &0.931& 1.26& 12.5& 6.82 & 5.92 & 4.2  \\
\raisebox{0.5\normalbaselineskip}[0pt][0pt]{B}
&BL& 0.1 & 0.01& 0.1& 0.2& $ 5 \times10^2$  &0.931& 1.26& 12.5& 6.82 & 5.95 & 4.2 \\
&CS& 0.01& 0.5 & 0.1& 0.01& $ 5 \times10^2$ &0.295& 63.1& 20.6& 5.94 & 5.35 & 14.4 \\
\raisebox{0.5\normalbaselineskip}[0pt][0pt]{C}
&CL& 0.01& 0.5 & 0.1&0.2& $ 5 \times10^2$   &0.295& 63.1& 20.6 & 5.94 & 5.38& 14.4 \\
&DS& 1   & 0.5 & 0.1& 0.01& $ 5 \times10^2$ &2.95 & 63.1& 28.7& 5.71 & 4.69 & 1.4 \\
\raisebox{0.5\normalbaselineskip}[0pt][0pt]{D}
&DL& 1   & 0.5 & 0.1&0.2& $ 5 \times10^2$   &2.95 & 63.1& 28.7& 5.71 & 4.69 & 1.4 \\
\hline
\end{tabular}
\end{center}
\end{table}

\clearpage

\onecolumn
\clearpage

\clearpage

\clearpage

\clearpage

\clearpage

\clearpage

\clearpage

\begin{figure}
\begin{center}
\includegraphics[width=160mm]{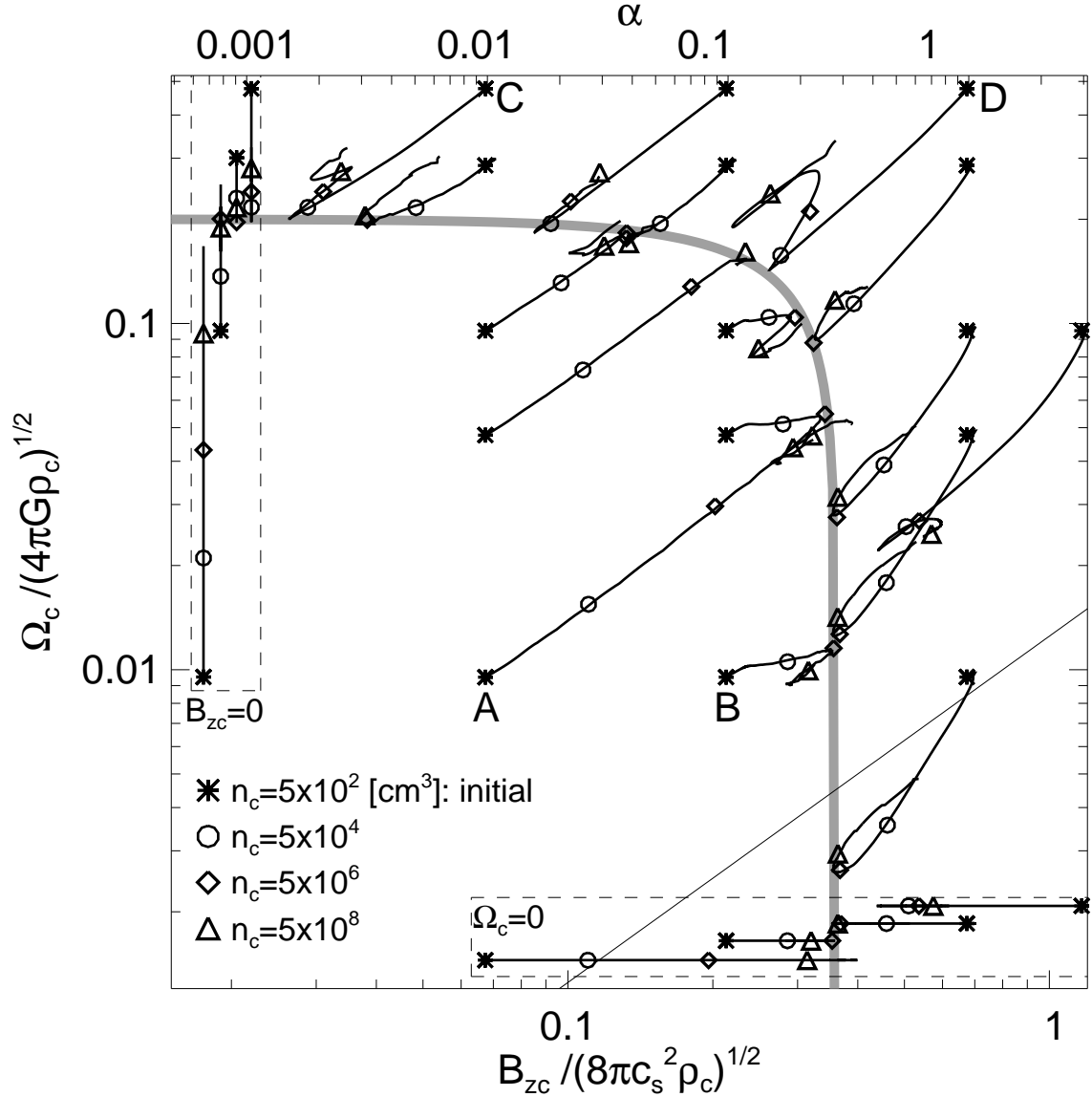}
\caption{
The evolutions of the magnetic flux density and angular velocity at cloud center.
The lower axis indicates the square root of the magnetic pressure ($\bzc/\sqrt{8 \pi}$) normalized by the square root of the thermal pressure ($\sqrt{ c_s^2 \rho_c}$) and the left ordinate does the angular speed ($\Omega_c$) normalized by the freefall timescale ($\sqrt{4 \pi G \rho_c}$), respectively.
The upper  axis indicates the parameter, $\alpha$.
The symbols, $*$, $\circ$, $\diamond$, and {\tiny $\triangle$}, represent at $n_c = 5 \times 10^2 \cm$ (initial state),  $5 \times 10^4 \cm$, $5 \times 10^6 \cm$, and $5 \times 10^8 \cm$, respectively.
Each line denotes the evolutional path from the initial state ($n_{\rm c,0} = 5\times 10^2 \cm$) to the end of the isothermal phase ($n_{\rm c} = 5 \times 10^{10} \cm$).
The characters 'A', 'B', 'C' and 'D' denote the model shown in Table~\ref{table:para}.
The thick curve denotes the magnetic flux and spin relation, $\Omega^2/[(0.2)^2 \times 4\pi G \rho_c] + B_{zc}^2/[(0.36)^2 \times 8\pi c_s^2 \rho_c] =1$ [see Equation~(\ref{eq:UL})].
The thin line indicates the relation of $\Omega_{\rm c}/(4 \pi G \rhoc)^{1/2} \propto B_{\rm zc}/(8 \pi c_s^2 \rhoc)^{1/2}$.
The models with no magnetic field are shown inside the upper left box, while those with no rotation are shown inside the lower right box.
}
\label{fig:U1}
\end{center}
\end{figure}
\clearpage

\end{document}